\documentclass[preprint,showpacs,preprintnumbers,amsmath,amssymb,superscriptaddress,aps,floatfix]{revtex4-1}

 \usepackage{graphicx}
 \usepackage{dcolumn}
 \usepackage{bm}

\begin{document}
\title{Precise Description of Single and Double Ionization of Hydrogen Molecule in Intense Laser Pulses} 

 \author{Mohsen Vafaee}
  \email{mo$\_$vafaee@sbu.ac.ir}
   \affiliation{Laser and Plasma Research Institute, Shahid Beheshti University, G. C., Evin, Tehran 19839-63113, IR Iran}
 \author{Firoozeh Sami}
   \affiliation{Laser and Plasma Research Institute, Shahid Beheshti University, G. C., Evin, Tehran 19839-63113, IR Iran}
 \author{Babak Shokri}
   \affiliation{Laser and Plasma Research Institute, Shahid Beheshti University, G. C., Evin, Tehran 19839-63113, IR Iran}
 \author{Behnaz Buzari}
   \affiliation{Department of Chemistry, University of Isfahan, Isfahan 81746-73441, IR Iran}
 \author{Hassan Sabzyan}
   \affiliation{Department of Chemistry, University of Isfahan, Isfahan 81746-73441, IR Iran}

\newcommand{\be}{\begin{equation}}
\newcommand{\ee}{\end{equation}}
\newcommand{\bea}{\begin{eqnarray}}
\newcommand{\eea}{\end{eqnarray}}
\newcommand{\h}{\hspace{0.30 cm}}
\newcommand{\vs}{\vspace{0.30 cm}}
\newcommand{\n}{\nonumber}
\begin{abstract}
A new simulation box setup is introduced for the precise description of the wavepacket evolution of two electronic systems in intense laser pulses. 
In this box, the regions of the hydrogen molecule H$_{2} $, and singly and doubly ionized species, H$_{2}^+ $ and H$_{2}^{+2} $, are well recognized and their time-dependent populations are calculated at different laser field intensities.
In addition, some new regions are introduced and characterized as quasi-double ionization and their time-dependencies on the laser field intensity are calculated and analyzed. 
The adopted simulation box setup is special in that it assures proper evaluation of the second ionization.
In this study, the dynamics of the electrons and nuclei of the hydrogen molecule are separated based on the adiabatic approximation. 
The time-dependent Schr\"{o}dinger and Newton equations are solved simultaneously for the electrons and the nuclei, respectively.
Laser pulses of 390 nm wavelength at four different intensities (i.e. $ 1\times10^{14} $, $ 5\times10^{14} $, $ 1\times10^{15} $, and $ 5\times10^{15} $ W cm$^{-2}$) are used in these simulations. 
Details of the central H$_{2} $ region is also presented and discussed. This region is divided into four sub-regions related to the ionic state H$^+$H$^-$ and covalent (natural) state HH. 
The effect of the motion of nuclei on the enhanced ionization is discussed.
Finally, some different time-dependent properties are calculated and their dependencies on the intensity of the laser pulse are studied, and their correlations with the populations of different regions are analyzed.

\end{abstract}

\maketitle
\section{Introduction}
Recently, the major focus of the theoretical and experimental research efforts  has been put
on the interaction of two electron systems with ultrashort intense laser pulses \cite{Dorner2002,Alnaser2004,McKenna2009,Manschwetus2009,Mauger2010,Becker2008,Liao2010,Chen2010,Guan2010,Camiolo,SSV1,Dehghanian2010,Ergler2005, Staudte-Chelkowski2007, Litvinyuk2008, Jin-Tong2010, Vafaee-8-10, Madsen_KER}.
This interaction causes many important phenomena such as single and double ionization, charge resonance enhanced ionization, dissociative-ionization, and high order harmonic generation \cite {review}.
Among these phenomena, single ionization has been intensively studied for many years. 
While, double ionization by intense laser fields has continued to draw considerable theoretical and experimental attention in the last two decades.
The double ionization can occur in two ways: i) each electron absorbs energy from the field independently (sequential) or ii) one electron absorbs the energy from the field and then shares it with the second electron via electron-electron correlation (nonsequential).
An important question is about the role of electron correlation in the double ionization. 
Despite its long history, the underlying question of the dynamics of electron correlation is still one of the fundamental puzzles in quantum physics \cite{Dorner2002,Chen2010,Liao2010}. 
 The first step for understanding the behaviour of many-electron systems in ultrashort intense laser pulse is to investigate the simplest two-electron systems such as the helium atom and the H$_{2}$ diatomic molecule.
One approach to describe this interaction is that only one electron is active and responsible for the interaction. This is called the single active electron (SAE) approximation. In this approximation, all other electrons are assumed to contribute in the dynamics of the active electron through a static screening potential. 
However, for the full description and understanding of the behaviour of two-electron systems, it is necessary to consider explicitly both electrons simultaneously. This plays a central role in developing our understanding of the interaction of two or many-electron atoms and molecules with ultrashort intense laser fields. 
For this, it is necessary to solve the time-dependent Schr\"{o}dinger equation (TDSE) for the two-electron systems 
that requires a huge amount of computations
which is beyond the capabilities of current computing facilities.
One way to overcome this difficulty is to reduce dimensions of the problem, accomponied by the use of soft-core potential \cite {Eberly1990, Kulander1996, Kstner2010, Camiolo, SSV1}.

In this work, we focus on the electrons dynamics in hydrogen molecule considering the nuclei as classical particles. This model reduces the complexity of the problem and helps detailing the electrons dynamics without involving the dissociative-ionization process\cite{Ergler2005, Staudte-Chelkowski2007, Litvinyuk2008, Jin-Tong2010, Vafaee-8-10, Madsen_KER}. The indistinguishability concept and the symmetry properties of the two electrons will be demonstrated with some details.
The paper is organized as follows: In Sec. II, details of the simulation box and the numerical solution of the TDSE are described. In Section III, results obtained for different intensities of the laser pulses for the introduced simulation box are presented and discussed. Finally, the conclusion appears in Sec.~IV.  
We use the atomic unit throughout this article unless stated explicitly.
\section{Numerical solution of the TDSE}
\begin{figure}[ht]
\centerline{\includegraphics[width=10cm]{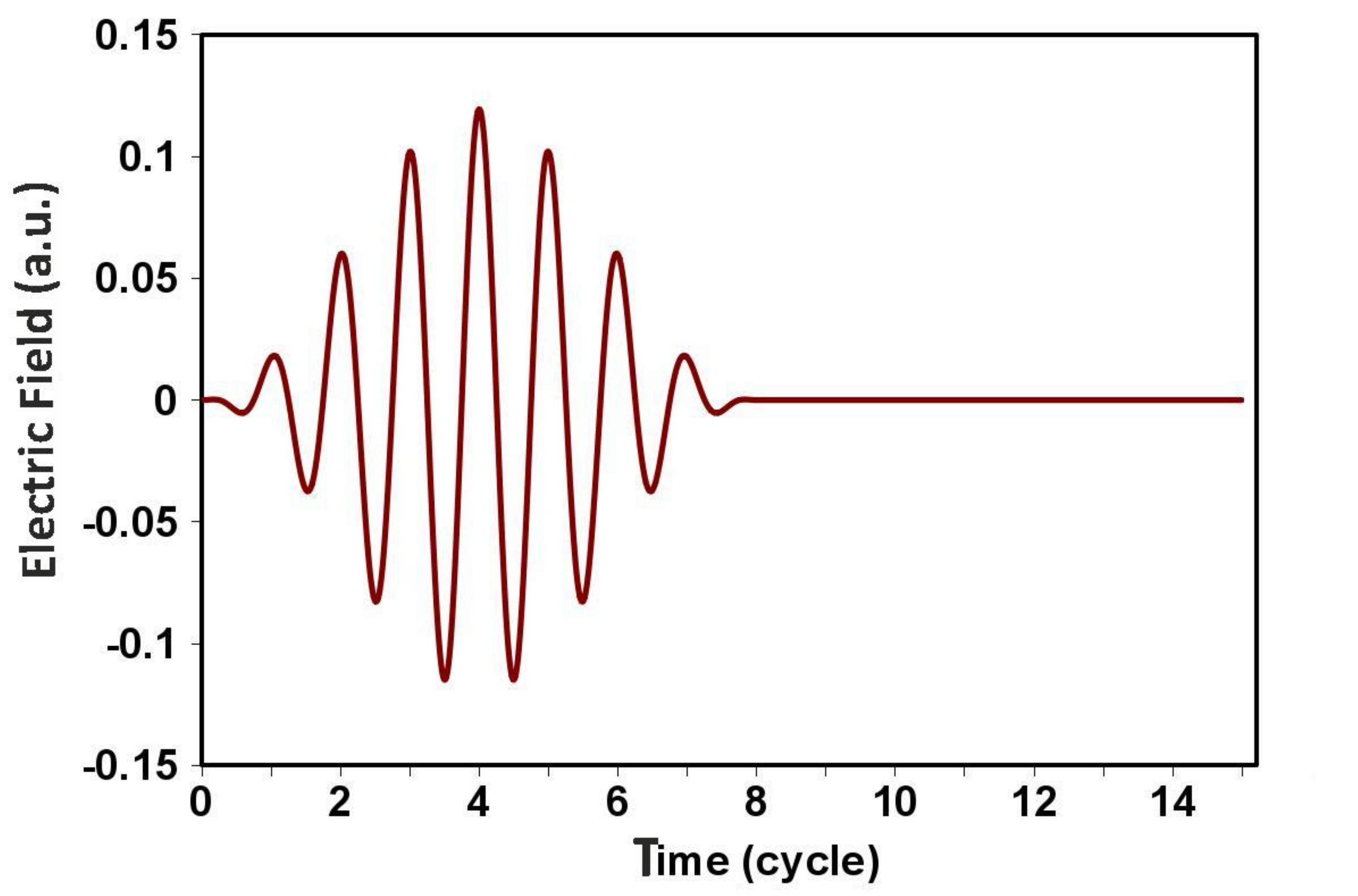}}
\caption{\label{field}\small The laser electric field used in this work has a sine squared envelop function with a duration of 8 optical cycles. }
\end{figure}

%
Assuming a linearly polarized laser pulse, and considering the fact that dynamics of the electrons and the nuclei occur in the laser field direction, we adopt a one-dimensional model for both the electrons and nuclei. In what follows, $R_1$ and $R_2$ indicate the nuclei positions and $z_1$ and $z_2$ are the electrons coordinates. Furthermore, $M$ and $m$ indicate the nucleus and electron masses, respectively, and $e$ is the electron charge. The temporal evolution of electrons of this system is described by the time dependent Schr\"{o}dinger equation (TDSE), i.e. \cite{Rigamonti,Camiolo,SSV1}
\begin{eqnarray}\label{eq:1}
i\frac{\partial \psi(z_1,z_2,t; R_1(t), R_2(t))}{\partial t}=
&&H_e(z_1,z_2,t; R_1(t), R_2(t)) \psi(z_1,z_2,t; R_1(t), R_2(t))
\end{eqnarray}
where the electronic Hamiltonian for this system, H$_e(z_1,z_2,t; R_1(t), R_2(t))$, is given by
\begin{eqnarray}
H_e(z_1,z_2,t; R_1(t), R_2(t)) =
&&-\frac{1}{2m_{e}} \left[\frac{\partial ^2}{\partial z_1^2}+\frac{\partial ^2}{\partial z_2^2}\right]\nonumber\\
&&+V_{C}(z_1,z_2,t; R_1(t), R_2(t)),
\label{eq:2}
\end{eqnarray}
\begin{eqnarray}
V_{C}(z_1,z_2,t; R_1(t), R_2(t)) =
&& \sum_{i, \alpha=1}^2 \left(\frac{-Z_\alpha}{\sqrt{( z_i-R_\alpha)^2+a }}\right)
 +\frac{1}{\sqrt{( z_1-z_2)^2+b }}+\frac{Z_1Z_2}{\sqrt{( R_1-R_2)^2+c }}\nonumber\\
&&+(z_1+z_2) E_0 f(t)\cos(wt),\nonumber\\
&&          
\label{eq:3}
\end{eqnarray}
where $ Z_1=1$ and $Z_2=1$ are the charges of nuclei and
the screening parameters $a$, $b$ and $c$ are responsible for the softening of the electron-nuclei, electron-electron and nucleus-nucleus interactions, respectively. The values of these parameters are set to the same values as used by Camiolo et al. \cite{Camiolo}.
The laser-molecule interaction energy is formulated in the dipole approximation, where $E_0$ is the laser peak amplitude and $\omega = 2\pi \nu $  is the angular frequency of the laser pulse. The laser pulse envelope, $f(t)$, is set as
\begin{eqnarray}
f(t) =sin(\dfrac{t}{\tau_{1}}\pi)^{2}
\label{eq:4}
\end{eqnarray}
where $ \tau_1 $ is the time duration of the field irradiation, set at $ \tau_1 $=8 cycles in this work. After this time, the simulation continues for 7 more cycles, i.e. with the laser field off, as shown in the Fig.~\ref{field}. 

The separated dynamics of the electrons and the nuclei are investigated in quantum and classic approaches, respectively. So, we solve the time-dependent Schr\"{o}dinger and Newton equations simultaneously for the electrons and the nuclei, respectively.
This implies that adiabatic approximation has been used to separate motions of electrons and nuclei \cite{Rigamonti}. 
The initial state is a singlet ground electronic state with an equilibrium internuclear distance $R=2.13$ at rest. 
In this singlet state, the fermion electrons adopt an antisymmetric spin configuration, and thus the electronic spatial part of the wavefunction $\psi(z_1,z_2,t; R_1(t), R_2(t))$ is symmetric with respect to the permutation of the two electrons.


\begin{figure}[ht]
	\resizebox{120mm}{!}{\includegraphics{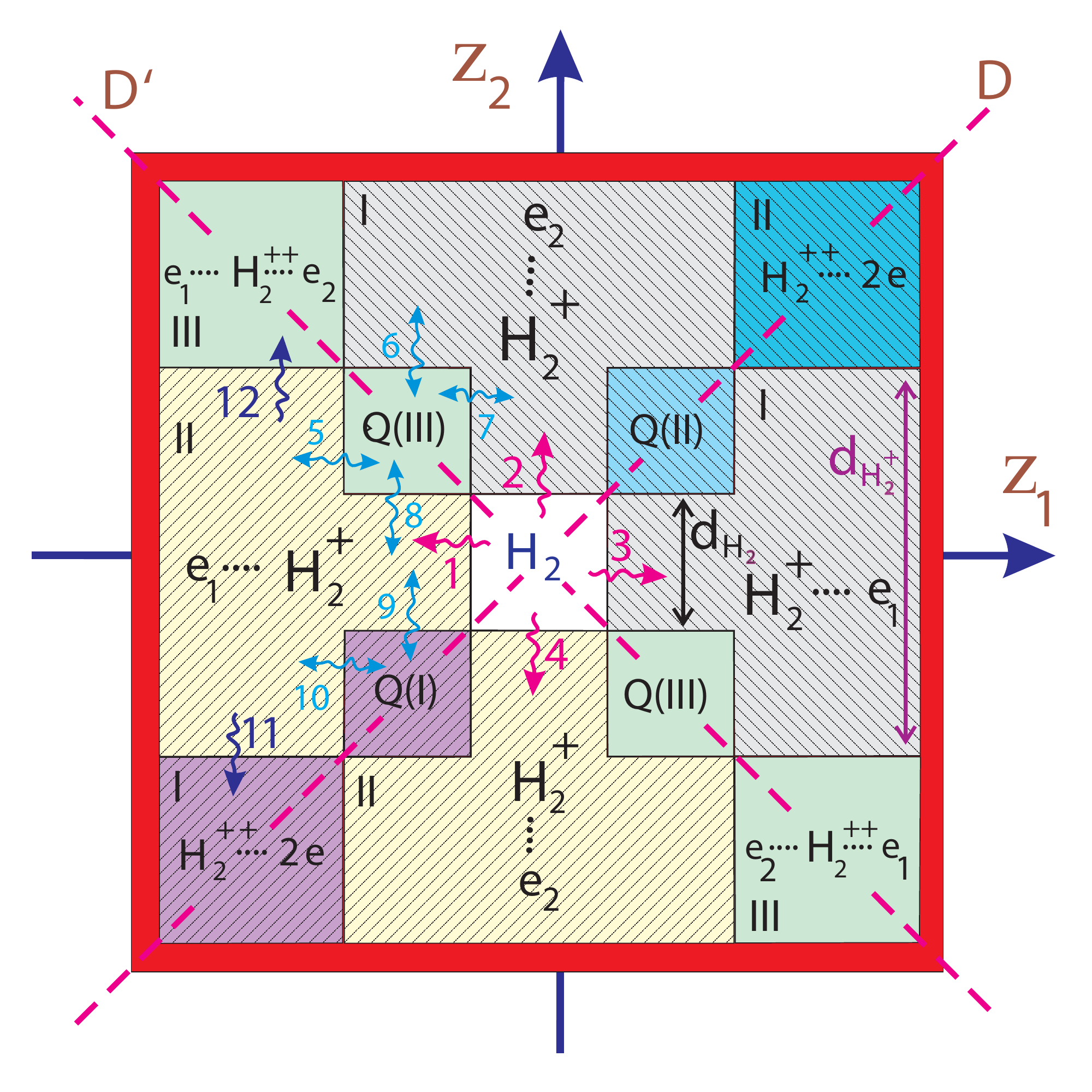}}
	  \caption
		{ 	\label{fig:f2} 
(Color online) The schematic diagram of the simulation box adopted in this study in which adjustable $d_{H_2}$ and $ d_{H_2^+}$ parametrs (with $d_{H_2^+} >  d_{H_2}$) are introduced.
		}
\end{figure}

\section{Simulation Box}
In our previous work, a box is introduced to be used for simulating the behaviour of H$_2$ in intense laser field \cite{SSV1}.
In the present study, we develope a new simulation box shown in Fig.~\ref{fig:f2}. 
Here also, the horizontal and vertical axes of the simulation box are assigned to the two $ z_1 $ and $ z_2 $ coordinates of the two electrons. Therefore, the $\psi(z_1,z_2,t; R_1(t), R_2(t))$ is symmetric with respect to the principle diagonal (D) of this coordinate system.
This symmetry decreases the processing time and the required amount of memory in the simulations, and therefore speeds up the calculations. 
The simulation is thus carried out only on the upper part of the box with respect to this D diagonal.
The initial state corresponds to the ground state of the hydrogen molecule, i.e. evolution of the system starts from the central H$_2$ region, Fig.~\ref{fig:f2}. 
When an electron of a molecule or ion moves a distance far enough away from the nuclei, it becomes ionized. 
This distance  is smaller for H$_2$ ($ d_{H_2}$) than that for H$_2^+$ ($ d_{H_2^+}$) because the electron of H$_2$ must escape from a core with one positive charge, whereas the electron of H$^{+}_{2}$ must escape from a core with two positive charges. 
To our knowledge, in all previous studies, see for example Refs. \cite{Camiolo,Saugout2008,ban-lu2005}, these two distances ($ d_{H_2}$ and $ d_{H_2^+}$) have been assumed to be equal.
In this work, we consider them to be different (Fig.~\ref{fig:f2}). 
This results in the appearance of some new geometries and regions which introduce interesting evolution stages for the wavepacket.

The first ionization of H$_2$ may occur via four pathways, denoted by arrows labled 1-4 in Fig.~\ref{fig:f2}. 
As stated in \cite{SSV1}, because of the diagonal symmetry with respect to the D line, pathways 3 and 4 correspond identically to pathways 1 and 2, respectively. 
In pathway 1, electron $e_1$ moves away from the nuclei in the negative direction, and in the pathway 2, electron $e_2$ gets distance from the nuclei in the positive direction. 
Therefore, there are two distinguishable regions which are labelled by H$_{2}^{+}$(I) and H$_{2}^{+}$(II) (Fig.~\ref{fig:f2}).

It is necessary to mention that beyond the H$_2$ region, the behaviour of the ionized electron is controlled by the laser electric field.
As it is seen in Fig.~\ref{fig:f2}, with introduction of $ d_{H_2^+} > d_{H_2}$, the regions H$_2^+ $(I) and H$_2^+  $(II) overlap in some spaces, labled as Q(I), Q(II) and Q(III) regions. 
In these overlapping regions, both $e_1$ and $e_2$ electrons are ionized. So it seems that these regions belong to H$_2^{+2}$. However, when $e_1$ or $e_2$ belonging to these regions, moves away, it enters the corresponding H$_2^{+}$ regions. For example, when $e_1$ belonging to the upper Q(III) region, leaves this region via pathway 5, $e_2$ is captured by the core unexpectedly, and as a result, the system enters the upper part of H$_2^{+}$(II).
This phenomenon can be explained as fallows: When the system is in the overlapping Q(I), Q(II) and Q(III) regions, the $e_1$  and $e_2$ electrons shield each other partially with respect to the charges of nuclei. 
When $e_1$ takes distance from the nuclei via pathway 5, its corresponding shielding effect on the $e_2$ electron is reduced. At this time, the $e_2$ electron which has come now closer to the nuclei, introduces more effective shielding on the $e_1$ electron so that the $e_1$ electron is sought to be ionized faster while the $e_2$ electron is pulled closer by the nuclei and thus enters the H$_2^{+2}$(II) region.
Similarly, in the upper Q(III) region, when the $e_1$ becomes closer to the nuclei (via pathway 7), it introduces a more effective shielding on the charges of the nuclei, and thus, the $e_2$ which is now farther from the nuclei, is less affected by the charges of the nuclei. Therefore, the $e_2$ will be ionized and thus the $e_1$ will be retracted by H$_2^{+2}$, and the system settles in the upper H$_2^{+}$(I) region.
Because of these interesting features of the Q regions, we call them as quasi-H$_2^{+2}$ regions.
The wavefunction in the Q(I) and Q(II) regions are different due to the symmetry breaking intriduced by the presence of the linearly polarized laser field. 
Therefore, we have three distinct quasi-H$_2^{+2}$ regions. 

In the Q(III) regions, the $e_1$ and $e_2$ electrons move away from origin in opposite directions. 
For example, in the upper Q(III) regions, $e_1$ is located in the negative part of the $z_1$ coordinate and $e_2$ is located in the positive part of the $z_2$ coordinate. 
For the quasi-H$_2^{+2}$ species in the Q(I) and the Q(II) regions, $e_1$ and $e_2$ are located in the same directions with respect to the nuclei, i.e.
in the Q(I) and Q(II) regions, $e_1$ and $e_2$ are both located respectively in the negative and positive parts of the $z_1$ and $z_2$ coordinates.

Since, all regions of the simulation box are symmetric with respect to the diagonal D, it is just necessary to solve the TDSE and calculate the wavefunction for one side of this diagonal; the results for the other side can be generated just by switching the $e_1$ and $e_2$ coordinates in the wavefunction.
For example, in the upper H$_2^{+}$(I) region, the behaviours of the $e_1$ and $e_2$ electrons are distinguishable, because $e_1$ is ionized and $e_2$ is bound.
The same is true for lower the H$_2^{+}$(I) region, but in which $e_2$ is ionized and $e_1$ is bound.
However, when both upper and lower H$_2^{+}$(I) regions are considered together, these distinguishability vanishes.
The same trends applies to other regions of the simulation box.
Therefore, distinguishability of the two electrons vanishes if both parts of the simulation box are considered.

If the intensity of the laser pulse after the first ionization is still strong enough to create the second ionization, the H$_2^{+}$ species can be ionized and the electrons are introduced to the H$_2^{+2}$ (I-III) regions. 
In this situation, it is necessary to follow and conserve H$_2^{+}$ in the simulation box till there is some probability for the second ionization, i.e. the H$_2^{+}$ regions must be extended so that H$_2^{+}$ is not absorbed by the boundary absorption.
This condition is not met in the setup of the simulation box in Fig. \ref{fig:f2} in which the H$_2^{+}$ may be absorbed by its absorbing boundaries and considered as H$_2^{+}$ species, before the laser pulse is turned off, and so, some chances of the second ionization is lost. 
Therefore, the rate of the second ionization in the simulation box of Fig.~\ref{fig:f2} may be underestimated. 
\begin{figure}
	\resizebox{120mm}{!}{\includegraphics{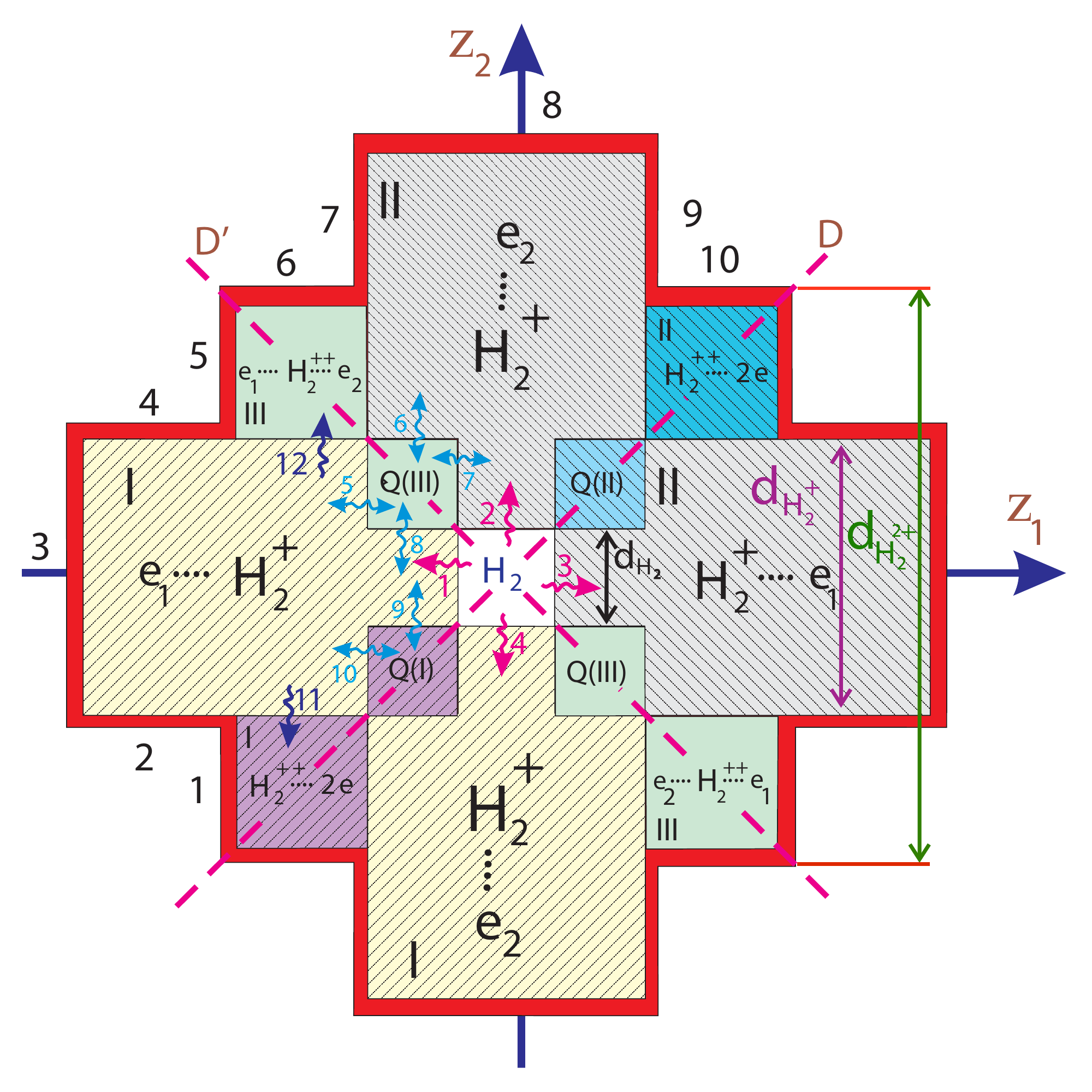}}
	  \caption
		{ 	\label{fig:f3} 
(Color online)  Same as Fig.~\ref{fig:f2}, but the H$_2^+$ regions are extended to assure that the overall second ionization is taken into account.
		}
\end{figure}
To cure this problem,
a new simulation box (Fig. \ref{fig:f3}) is introduced in which the H$_2^{+}$ components is conserved 
until the laser pulse is turned off to assure that the overall second ionization is taken into account completely. 
The H$_2^{+}$ regions in this simulation box (Fig.~\ref{fig:f3}) are extended farther along both directions  of $z_1$ and $z_2$  in comparison with Fig.~\ref{fig:f2}. 
In this setup of the simulation box, when an electron evolves from the H$_2^{+}$ regions to the H$_2^{+2}$ regions, 
or it is absorbed by the borders labelled 2, 4, 7, and 9, the second ionization takes place. So, the overall time-dependent second ionization is equal to the population in the H$_2^{+2}$ regions plus the population absorbed by the 1, 2, 4-7,9 and 10 absorbing borders in the upper part of the box and their corresponding populations in the lower part of the box.
Therefore, the overall time-dependent population of H$_2^{+}$ is equal to the population existing in the H$_2^{+}$ regions plus the population absorbed by the absorbing borders 3 and 8 and their corresponding borders in the lower part of the box. 

%
We present the results of simulation for two different cases of fixed and freed nuclei in the latter of which the dynamics of the nuclei is treated classically.
Laser pulses of 390 nm wavelength with four different intensities of $ 1\times10^{14} $, $ 5\times10^{14} $, $ 1\times10^{15} $, and $ 5\times10^{15} $ W cm$^{-2}$ are used. 
The shape of the laser pulse is shown in Fig.~\ref{field}.
The time step is set to $ \delta t$= 0.02. The differential operators in Eq.~(2) are discretized by the 11-point difference formulas which have tenth-order accuracies \cite{vafa2006}. To solve the above TDSE numerically, we adopted a general nonlinear coordinate transformation for electronic coordinates. For the spatial discretization, we have constructed a finite difference scheme with a nonuniform (adaptive) grid for the $z_1$ and $z_2$ electronic coordinates, which are finest near the nuclei and coarsest at the border regions of the simulation box. 
The size of the computational box along both $z_1$ and $z_2$ axes is $\pm785 a.u.$ and the width of the absorption regions are $\pm30 a.u.$. For each of $z_1$ and $z_2$ axes, 4000 points are considered . The interval between these points is 0.2 a.u. near the nuclei and becomes $\sim0.75 a.u.$ at the border regions of the simulation box. 
The absorber regions are introduced by using fourth-order optical potentials at the $z_1$ and $z_2$ boundaries, in order to capture the photoelectrons and prevent the reflection of the outgoing wave packets at the borders of the grid. More details of our calculations are described in our previous work \cite{vafa2004}. 
Our simulations for different box sizes and grid points show that the larger box sizes and more dense grid points does not improve the results. 
More details of our calculations are described in our previous reports \cite{vafa2006,vafa2004}.

\section*{Simulation results}
\begin{figure}
\centerline{\includegraphics[width=10cm]{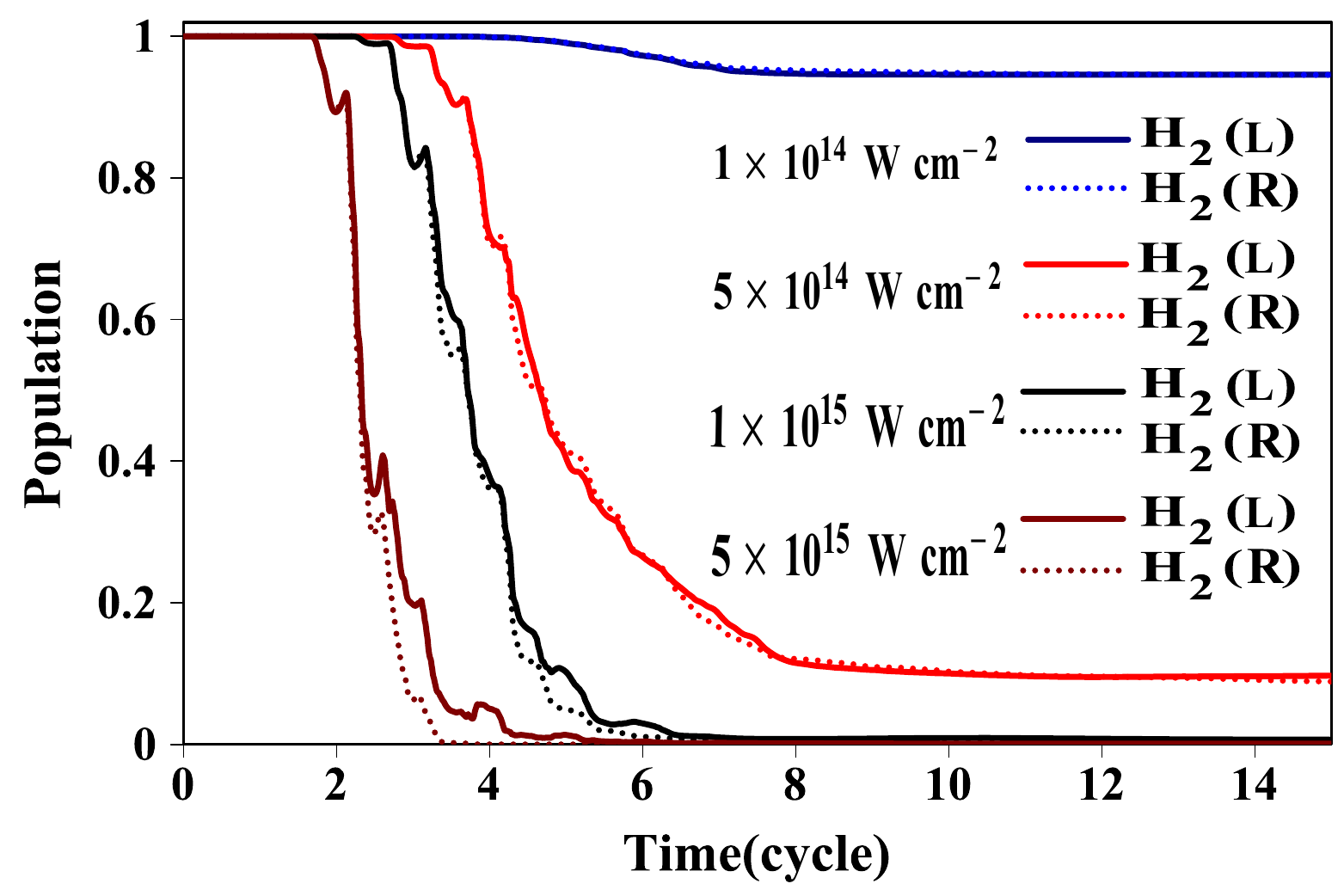}}
\caption{\label{DH2S}
\small (Color online) Time-dependent population of the H$_{2}$ region for the limited (L) and full (F) simulation boxes obtained at different laser pulse intensities.}
\end{figure}

\begin{figure}
\centerline{\includegraphics[width=10cm]{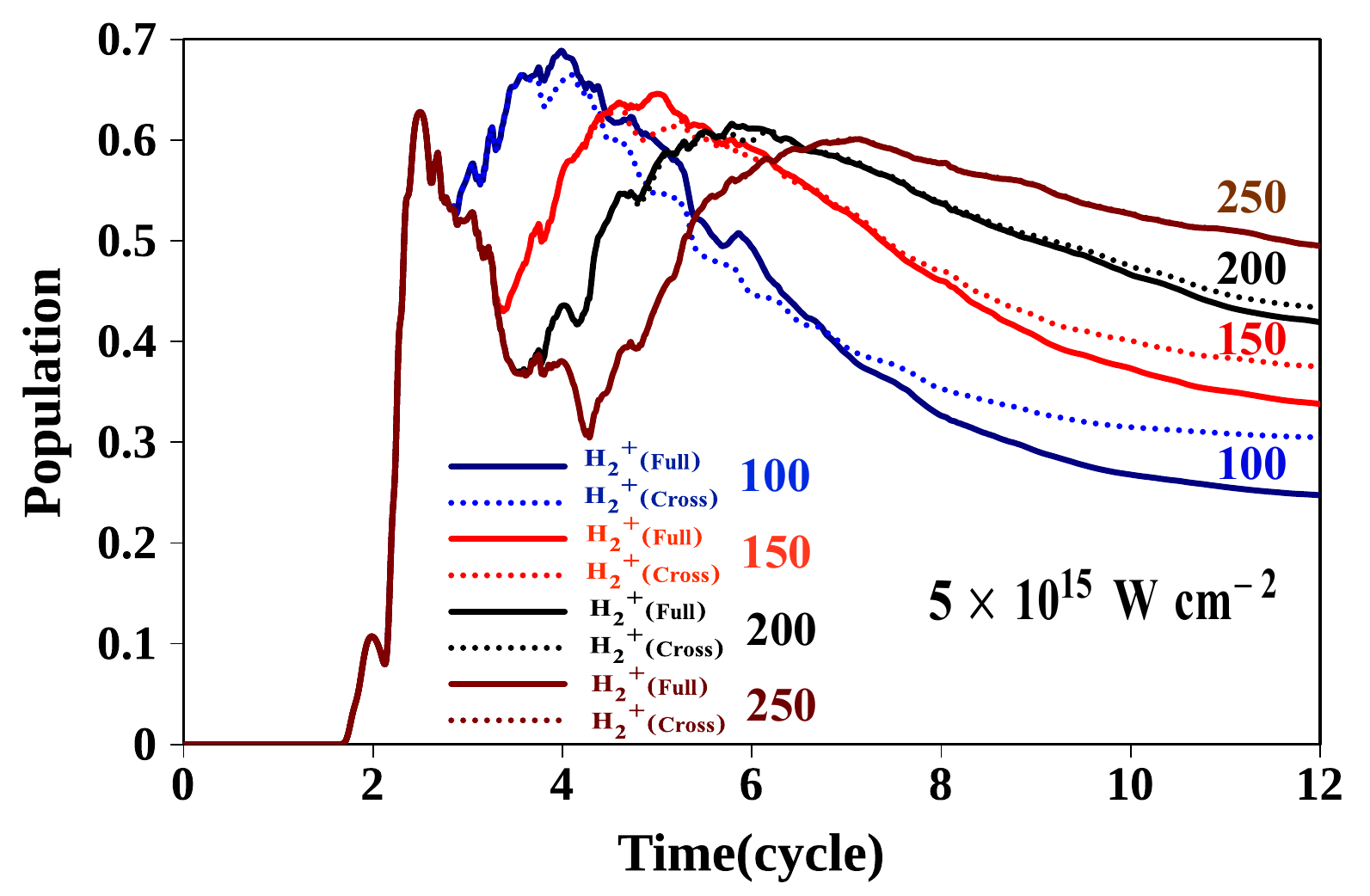}}
\caption{\label{DH2+S}
\small (Color online)
Time-dependent population of the $H_2^+$ region for different simulation boxes, respectively with
$d_{H_2^+}=200$ ($d_{H_2^2+}=280$ (cross) and $1510$ (full)),
$d_{H_2^+}=300$ ($d_{H_2^2+}=380$ (cross) and $1510$ (full)),
$d_{H_2^+}=400$ ($d_{H_2^2+}=480$ (cross) and $1510$ (full)), and
$d_{H_2^+}=500$ ($d_{H_2^2+}=580$ (cross) and $1510$ (full)), at 
laser intensity of $ 5\times10^{15} $ W cm$^{-2}$.
} 
\end{figure}

To determine the optimum values of the $d_{H_2}$, $d_{H_2^+}$, $d_{H_2^2+}$ geometric parameters of the simulation box (Fig.~\ref{fig:f3}), results obtained for simulation boxes with different values of these parameters are evaluated and compared. 
Fig.~\ref{DH2S} shows the time-dependent population of the H$_{2}$ region for two different simulation boxes with $d_{H_2}=60 a.u.$ once with a box limited (L) to the H$_{2}$ region only with no other regions, and once with the full box (F) as shown in Fig.~\ref{fig:f2}.
It can be seen from Fig.~\ref{DH2S} that variations of the populations for the L and F boxes are almost similar. 
For the comparative study intended here, a fixed size of $d_{H_2}=60$ for the H$_{2}$ region suffices. However, for precise evaluation of the population transfer between regions, this size should be optimized at each intensity. 
%
To adjust the $d_{H_2^+}$ and $d_{H_2^{2+}}$ values, population of the H$_2^+$ region obtained with different values of these parameters are calculated and compared for the cross box (Fig.~\ref{fig:f3}) and the full box (Fig.~\ref{fig:f2}) and plotted in Fig.~\ref{DH2+S}. This figure 
shows that, at the laser intensity of $5\times10^{15}$ W cm$^{-2}$, population of H$_2^+$ in the cross box with $d_{H_2^+}=500$ and $d_{H_2^2+}=580$ is converged to that of the full box (with $d_{H_2^+}=500$ and $d_{H_2^2+}=1510$).
The same results are obtained for other laser intensities examined in this work.
Therefore, for all intensities studied in this work, we set $d_{H_2}$, $d_{H_2^+}$ and $d_{H_2^2+}$ sizes at 60, 500 and 580 (i.e. corresponding to the positions of the region boundaries at $\pm30$, $\pm250$ and $\pm290$), respectively. 

\section*{H$_{2} $ regions}

\begin{figure}[h]
	\resizebox{100mm}{!}{\includegraphics{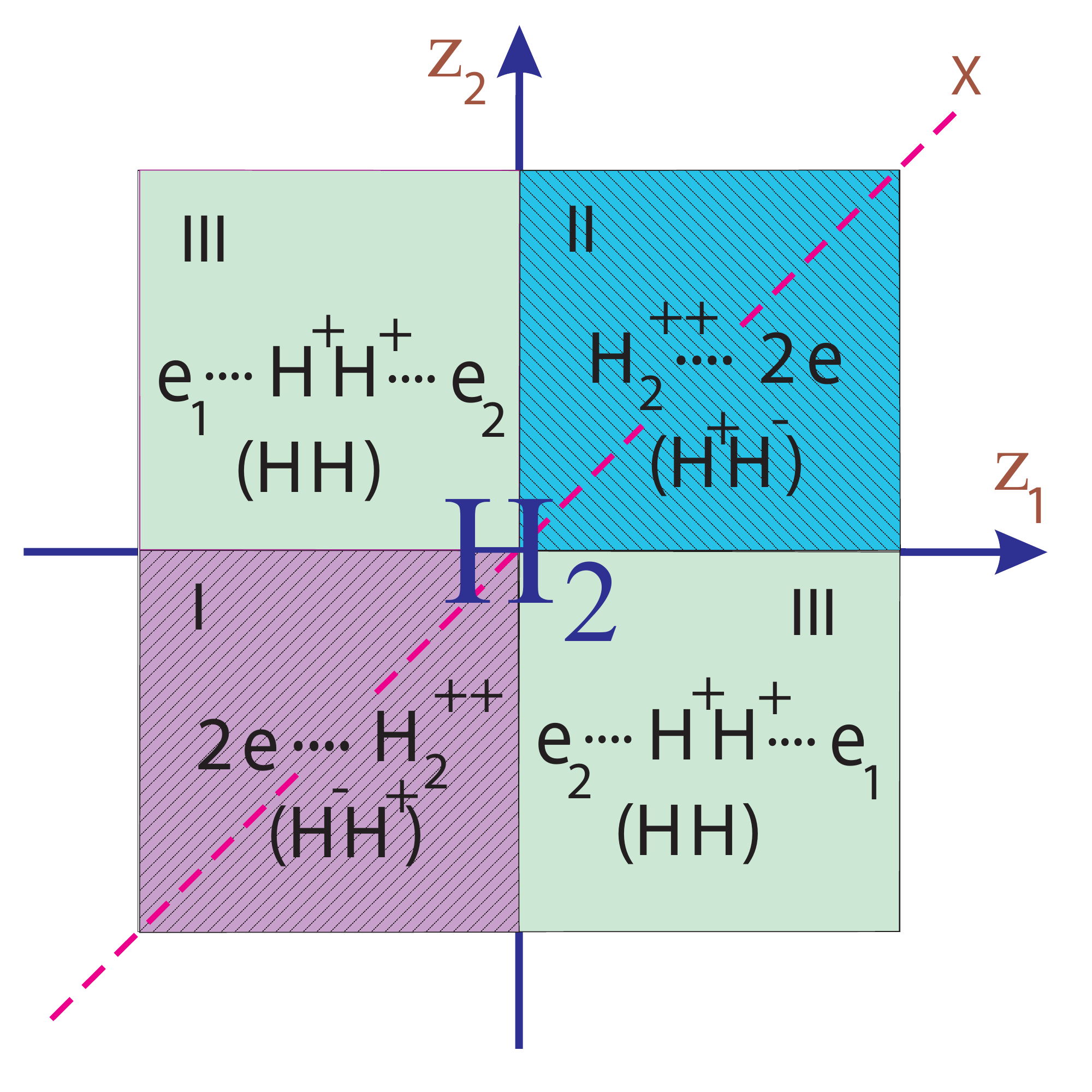}}
	  \caption
		{ 	\label{fig:H2_region} 
(Color online)  Details of the H$_2$ region of the simulation box introduced in Fig.~\ref{fig:f3}.
		}
\end{figure}

\begin{figure}
\centerline{\includegraphics[width=16cm]{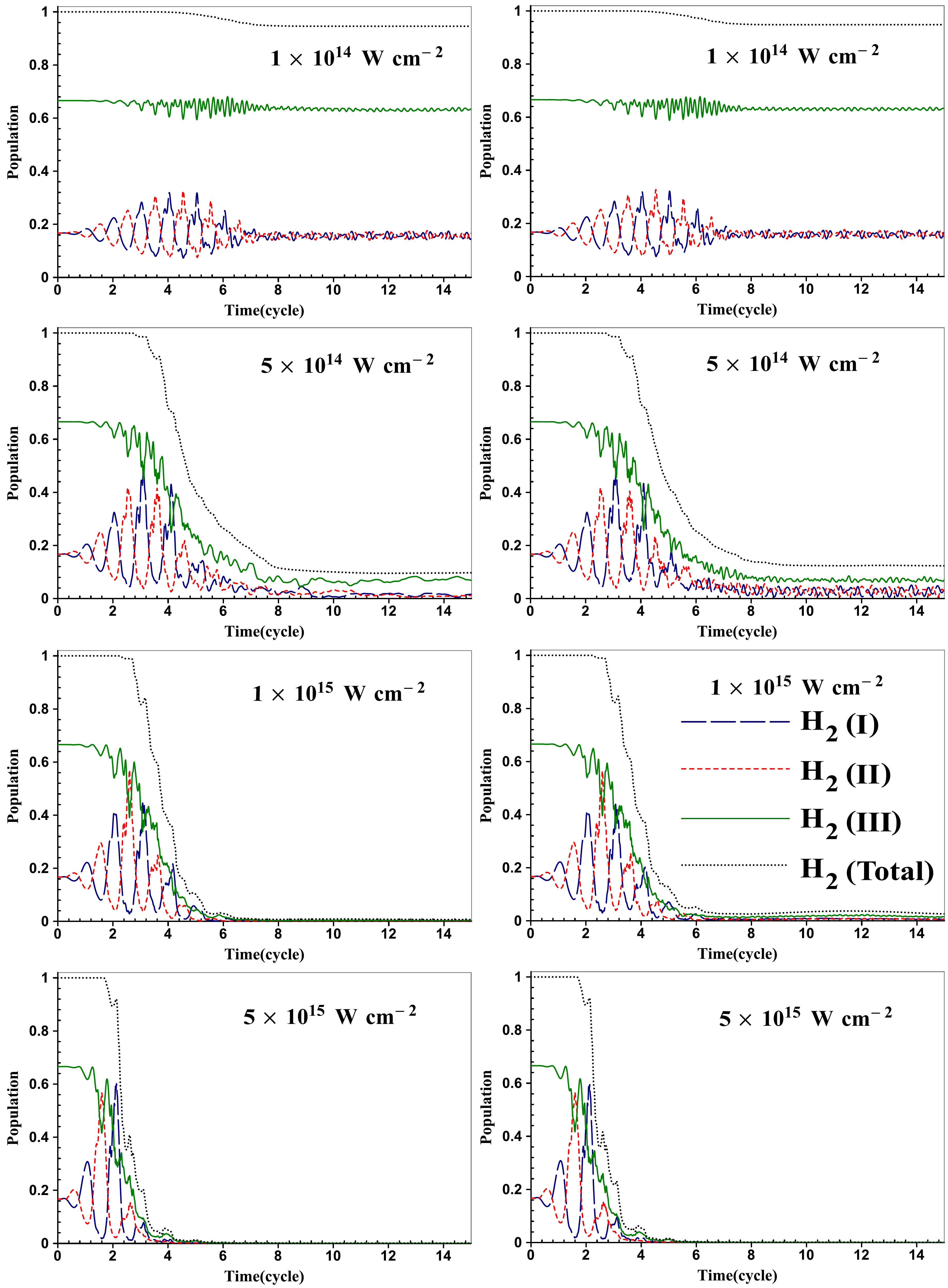}}
\caption{\label{H2}
\small (Color online) Time-dependent populations of different H$_{2}$ regions. The left and right sets of panels are related to the simulations with freed and fixed nuclei, respectively.}
\end{figure}

As shown in Fig.~\ref{fig:H2_region}, the H$_{2} $ region can be divided into four sub-regions. The H$_{2}$(I) and (II) regions are related to the ionic state H$^+$H$^-$, while the H$_{2}$(III) regions are related to the covalent (natural) state HH. The time-dependent populations of these different regions obtained at different laser pulse intensities for both cases of fixed and freed nuclei are calculated and plotted in the Fig.~\ref{H2} .
It can be seen from this figure that, at the intensity of $ 1\times10^{14}  Wcm^{-2}$, the ionization is low and the outgoing population from the H$_{2} $ region is negligible.

Under the irradiation of the laser field, the initial symmetric relations between the H$_{2}$(I) and H$_{2}$(II) regions disappears and, as can be seen in Fig.~\ref{H2}, their electron populations vary independently.
In the midway of the passage of the exchange of population between the H$^-$H$^+$(I) and H$^+$H$^-$(II) ionic regions, populations of the covalent HH(III) regions increase.
This feature of the wavepacket evolution appears as a small increase in the population of the H$_{2}$(III) region in the time intervals that the populations of the two ionic regions cross, Fig.~\ref{H2}.
Lower populations of the H$_{2}$(I) and H$_{2}$(II) regions relative to that of the H$_{2}$(III) region in the absence of the laser field is obvious. 
At the initial stages of the laser pulse, the field is not strong enough to alter this relative order of the populations.  
As the laser intensity increases, the outgoing population from the H$_{2} $ region increases due to the single ionization process. 
The amplitudes of the oscillations of the populations of the H$_{2}$(I) and H$_{2}$(II) regions increase with increasing the intensity of the laser pulse and at some instances the populations of the H$_{2}$(I) and H$_{2}$(II) regions become more than that of the H$_{2}$(III) regions. 
Therefore, the high laser field can increase population of the unstable H$_{2}$(I) and H$_{2}$(II) regions over that of the more stable H$_{2}$(III) regions. 
Figure~\ref{H2} also shows that when the population of the H$^-$H$^+$(I) and H$^+$H$^-$(II) ionic regions take over that of the covalent HH(III) regions, the overall population of the full H$_{2}$ region decreases considerably due to the ionization.
It worth mentioning that Dehghanian et al. \cite{Dehghanian2010} estimated the critical internuclear  distance (R$_{C}=5 a.u.$) for the enhanced single ionization probability (EI) using an electrostatic model based on the creation of the precursor ionic state H$^+$H$^-$ from the covalent (natural) state HH occuring at the peak of the laser pulse, and showed that double-ionization probability has the same R$_{C}$ value as that of the single-ionization probability.
The results obtained with fixed and freed nuclei are very similar (the left and right sets of panels of Fig.~\ref{H2}). 
\section*{quasi$-H_{2}^{2+} $ regions}

\begin{figure}
\centerline{\includegraphics[width=16cm]{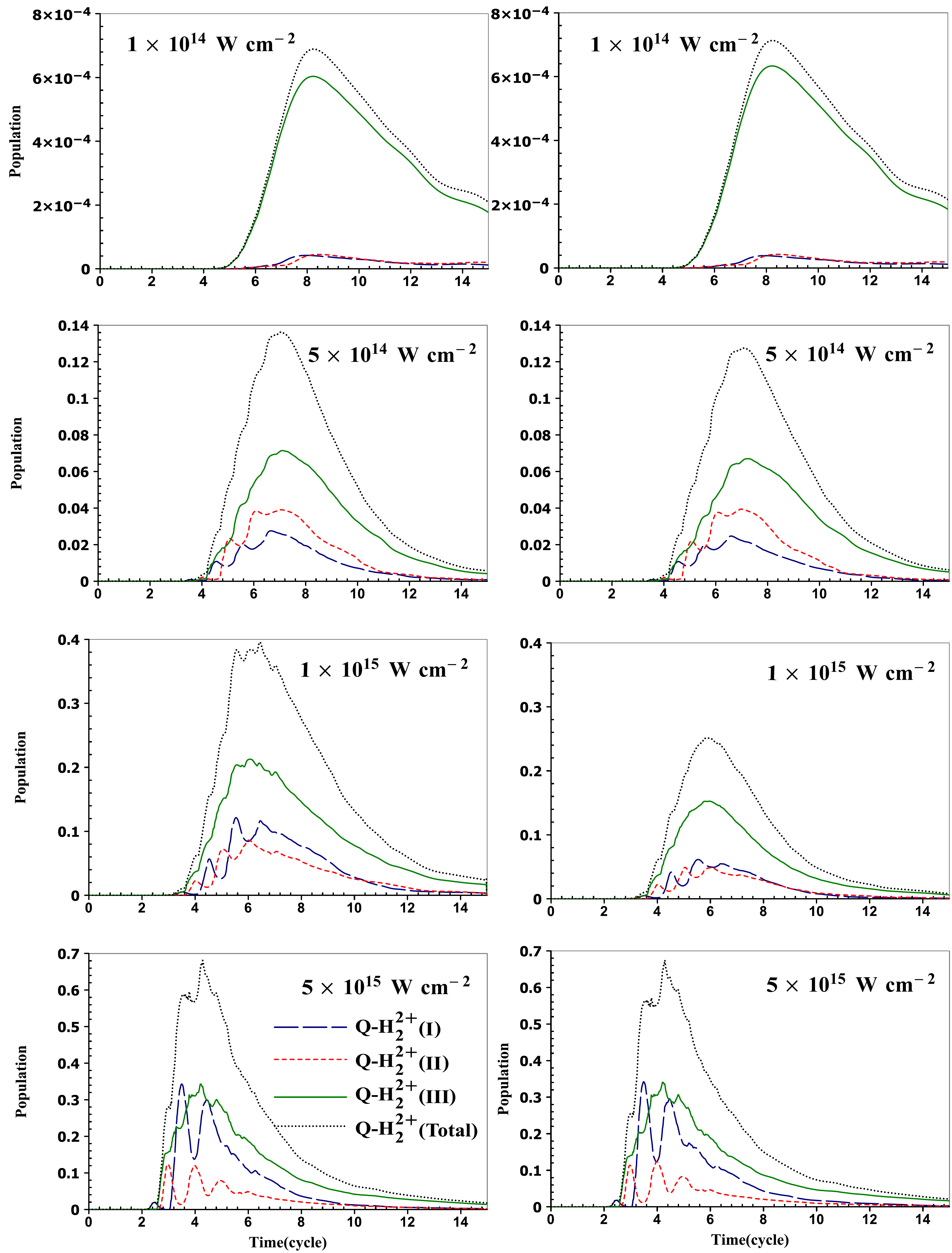}}
\caption{\label{1qh2++}
\small (Color online) The same as Fig.~\ref{H2}, but for the Q-H$_{2}^{2+}$ regions (Fig.~\ref{fig:f3}) for the case of freed (left) and fixed (right) nuclei. 
}
\end{figure}

The time-dependent populations of the Q-H$_{2}^{2+}$ regions are shown in Fig.~\ref{1qh2++}. At the intensity of $ 1\times10^{14}$  W cm$^{-2}$, there is only a small population in the Q-H$_{2}^{2+}$ regions most of which is in the Q-H$_{2}^{2+}$(III) region. 
For this intensity, the time-dependent behaviour of the population is the same for both cases of fixed and freed nuclei. As the laser intensity increases, the probability of the ionization increases and so the Q-H$_{2}^{2+}$ regions become more populated: the populations of the Q-H$_{2}^{2+}$(I) and Q-H$_{2}^{2+}$(II) regions are increased more as compared to that of the Q-H$_{2}^{2+}$(III) region.
In each half cycle of the laser pulse, a sharp rise and fall is observed for the population of the Q-H$_{2}^{2+}$(I) and the Q-H$_{2}^{2+}$(II) regions. 
While, the population variation of the Q-H$_{2}^{2+}$(III) region is smoother. 
When the laser field is turned off, due to the Coulomb repulsion force between the electrons, the Q-H$_{2}^{2+}$(I) and the Q-H$_{2}^{2+}$(II) regions are evacuated such that their population become negligible. 

\section*{H$_{2}^{+}$ regions}

\begin{figure}
\centerline{\includegraphics[width=16cm]{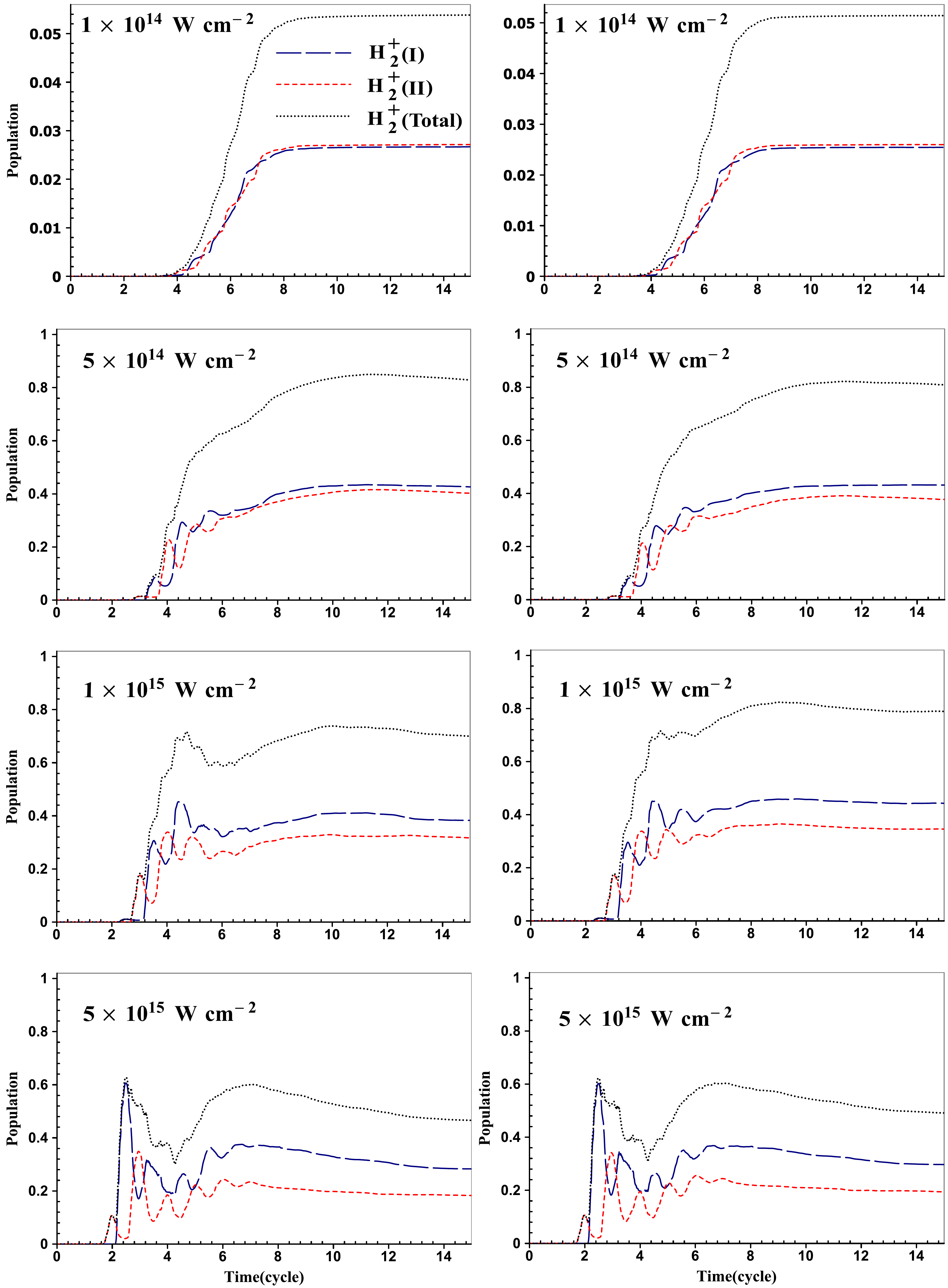}}
\caption{\label{1h2+}
\small (Color online) The same as Fig.~\ref{H2}, but for the singly-ionized regions H$_{2}^{+}$ (Fig.~\ref{fig:f3}) for the case of freed (left) and fixed (right) nuclei}
\end{figure}

Time-dependent populations of the singly-ionized regions H$_{2}^{+}$(I)  and H$_{2}^{+}$(II) (Fig.~\ref{fig:f3}) are shown in Fig.~\ref{1h2+}.
At the intensity of $ 1\times10^{14} $ W cm$^{-2}$, after four optical cycles, at the maximum amplitude of the laser pulse (Fig.~\ref{field}), the negligible population of the single ionization region starts to increase and then becomes constant until the laser field is turned off.
With increasing the laser intensity, populations of these regions increases drastically.
Fig.~\ref{1h2+} shows a correlation between populations of the H$_{2}^{+}$(I) and H$_{2}^{+}$ (II) regions which can be regarded as a population exchange via H$_{2}$ and Q-H$_{2}^{2+}$ intermediate regions.
Comparison of the corresponding panels of 
Figs.~\ref{H2}-\ref{1h2+} 
shows that this population exchange occurs mainly through the H$_{2}$ regions at low intensity and through the Q-H$_{2}^{2+}$ region at high intensity. 
This population exchange becomes more evident as the intensity of the laser field increases.
Also, the difference between the populations of the H$_{2}^{+}$(I) and H$_{2}^{+}$(II) regions increases with the laser intensity.
From Figs.~\ref{1h2+}, it is clear that at the intensity of $ 5\times10^{14}$ W cm$^{-2}$, the population of the singly-ionized H$_{2}^{+}$ regions is the highest compared to the other regions in Fig.~\ref{fig:f3}.
Except at $1\times10^{14}$ W cm$^{-2}$ intensity, population of the singly-ionized regions is higher than that of all other regions (Fig.~\ref{fig:f3}). 
The populations of the singly-ionized regions at the $ 1\times10^{15}$ and $ 5\times10^{15}$ W cm$^{-2}$ intensities show a peak in the time zones near the peak of the laser pulse and decreasing afterwards.
This decrease is due to the outgoing of the population from the singly-ionized to the doubly-ionized regions ones.

\section*{H$_{2}^{2+}$ regions}

\begin{figure}
\centerline{\includegraphics[width=16cm]{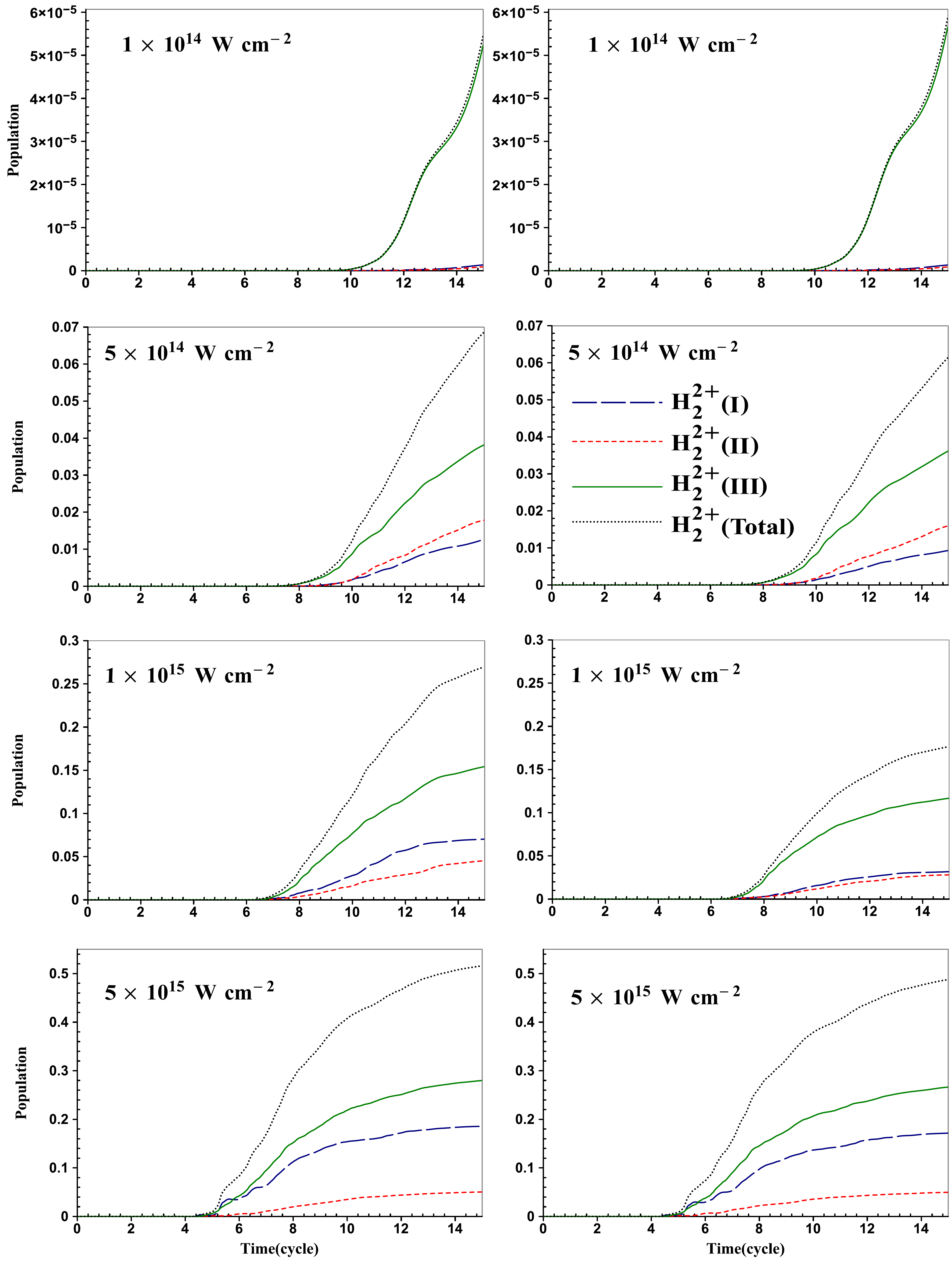}}
\caption{\label{1h2++tot}
\small (Color online) The same as Fig.~\ref{H2}, but for the doubly-ionized regions H$_{2}^{2+}$ (Fig.~\ref{fig:f3}) for the case of freed (left) and fixed (right) nuclei}
\end{figure}

Figure \ref{1h2++tot} shows the time-dependent populations of the doubly-ionized regions, H$_{2}^{2+}$(I), H$_{2}^{2+}$(II) and H$_{2}^{2+}$(III) (Fig.~\ref{fig:f2}). As can be seen from this figure, at the intensity of  $ 1\times10^{14}$ W cm$^{-2}$, populations of these regions are all negligible.  
As the laser intensity increases, these populations grow: the higher the intensity, 
the earlier the starting point of the sharp increase in the population.
Fig.~\ref{1h2++tot} shows that the rapid rise in the population of the doubly ionized species occurs near the peak of the laser pulse. 
The populations of both doubly ionized H$_{2}^{2+}$(I) and H$_{2}^{2+}$(II) regions are the same at low laser intensity. However, with increasing the laser intensity they become significantly different. 
Time-dependent population behaviours of different doubly-ionized regions are very similar for the both cases of fixed and freed nuclei
but their values are slightly different: populations of the latter is slightly higher.

\begin{figure} 
\centerline{\includegraphics[width=16cm]{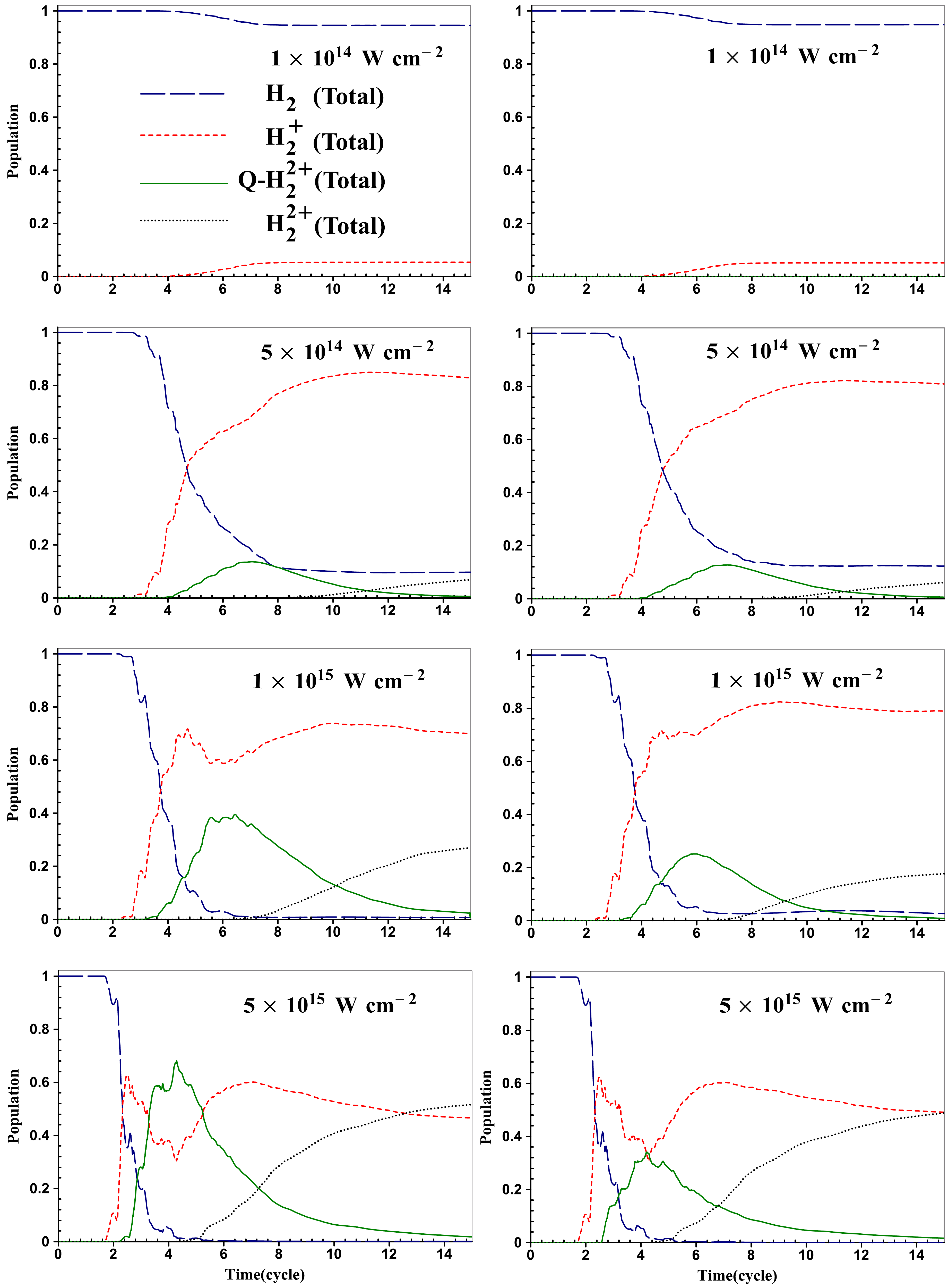}}
\caption{\label{totalnorm}
\small (Color online) The same as Fig.~\ref{H2}, but for comparison of overall populations of the H$_{2}$, H$_{2}^{+}$, Q-H$_{2}^{2+}$, and H$_{2}^{2+}$ regions.
The same line styles is used for different species in all panels.
}
\end{figure}

Variation of the overall populations of the H$_{2}$, H$_{2}^{+}$, Q-H$_{2}^{2+}$, and H$_{2}^{2+}$ regions are shown in Fig.~\ref{totalnorm}. As can be seen from this figure, at the lowest intensity of $ 1\times10^{14}$ W cm$^{-2}$, H$_{2}$ has the dominant population, while at the intensities of $ 5\times10^{14}$ and $ 1\times10^{15}$ W cm$^{-2}$ H$_{2}^{+}$ has the dominant population. 
At the highest intensity, $ 5\times10^{15}$ W cm$^{-2}$, population of the doubly ionized state H$_{2}^{2+}$ is dominant.

\section*{Position expectation value of electrons}

\begin{figure}
\centerline{\includegraphics[width=15cm]{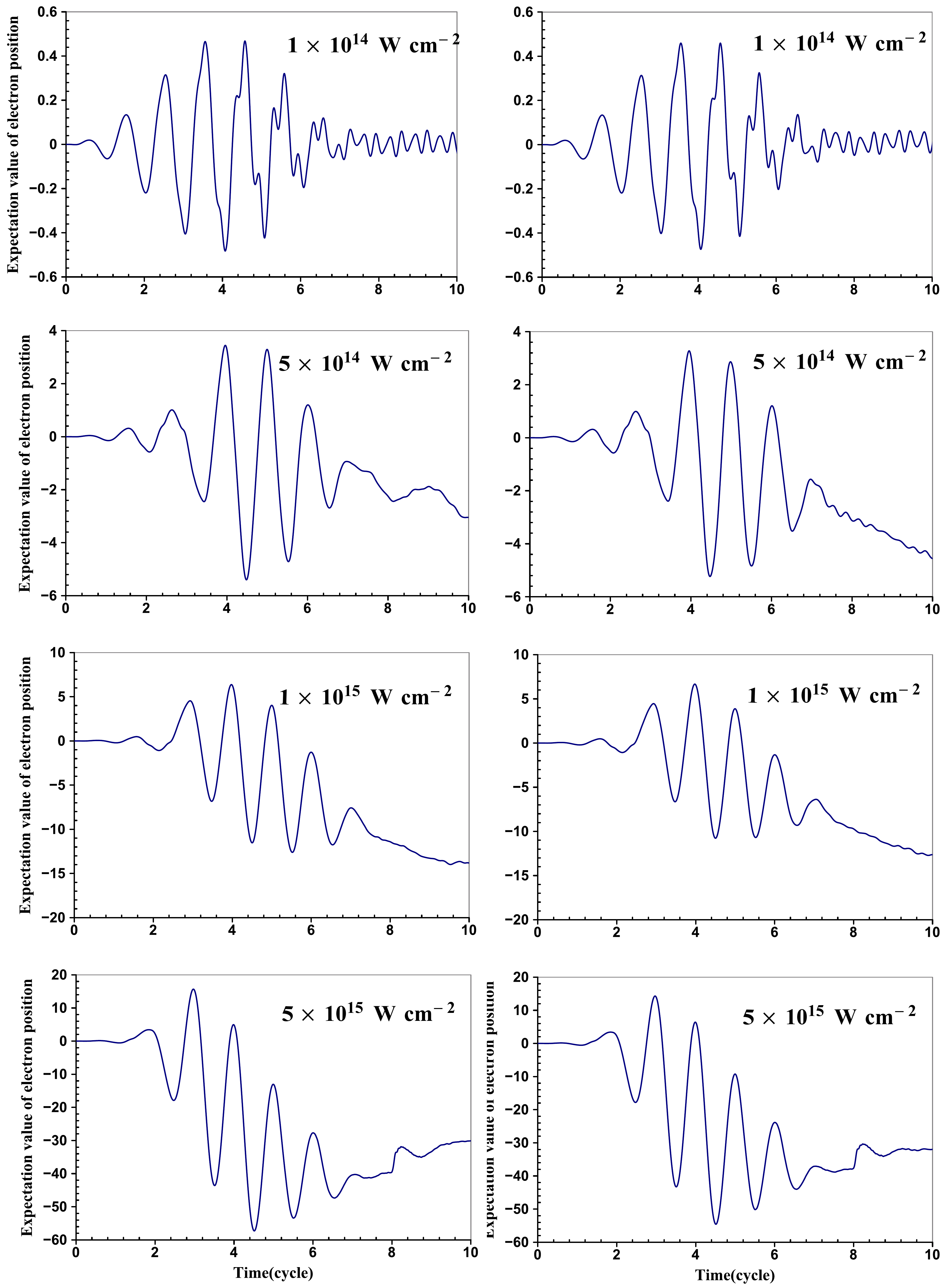}}
\caption{\label{1zvalue}
\small (Color online) Variation of the position expectation value of electrons.
The left and right sets of panels are related to the simulations with freed and fixed nuclei, respectively.
}
\end{figure}

Figure~\ref{1zvalue} shows the expectation (average) value of the position of the electrons calculated for two cases of fixed and freed nuclei.
For both cases, this expectation value show large-amplitude oscillations. 
When the electrons under the effect of the laser field are withdrawn in one side from the nuclei, the attraction influence of the nuclei on the electrons decreases. By changing the laser field direction, the electrons come toward the nuclei and then pass over it in the opposite direction. This results in the oscillation of the position expectation value in each cycle of the laser field. 
At the intensity of $ 1\times10^{14}$ W cm$^{-2}$, these oscillations loose their amplitude and become complex after turning off the laser field which ise due to the fact that the electronic wave function is now a linear superposition of different stationary states. 

At the intensity of $ 5\times10^{14}$ W cm$^{-2}$, after the laser field termination, the resulting electronic wavepacket recedes from the nuclei.
In this condition, some parts of the wave function go far away from the nuclei and result in an increase in the expectation value of position. 
Very small population transfer from H$_{2}^{+}$ to H$_{2}^{2+}$ in Fig.~\ref{totalnorm} at the intensity of $ 5\times10^{14}$ W cm$^{-2}$ after 8 cycles, explains this event. 
At the intensity of $ 1\times10^{15}$ W cm$^{-2}$, the population transfer are considerable such that population transfer from H$_{2}^{+}$ to H$_{2}^{2+}$ starts earlier, about 7 cycles (Fig.~\ref{totalnorm}).

At the highest intensity of $ 5\times10^{15}$ W cm$^{-2}$, a considerable population of H$_{2}$ is doubly ionized (i.e. the electron goes out of the simulation box). In this case, with the population transfer from  H$_{2}^{+}$ to H$_{2}^{2+}$, the reduction of the position expectation value happens.

\section*{Motion of Nuclei}

\begin{figure} [ht]
\centerline{\includegraphics[width=13cm]{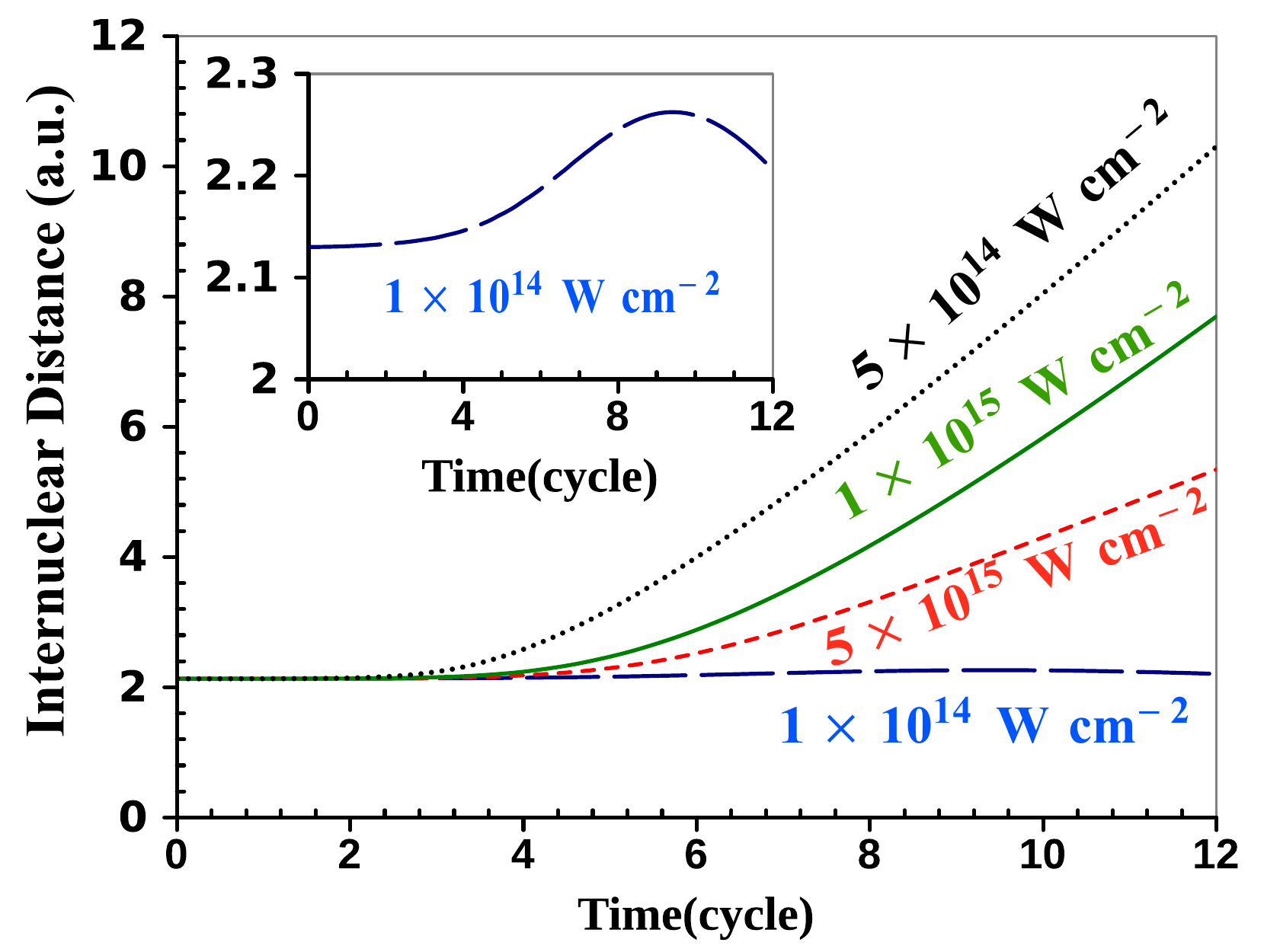}}
\caption{\label{nucdisfor}
\small (Color online) Time-dependent H-H internuclear distance in the interaction of two electrons 1D H$_{2}$ system with laser field of different intensities.
}
\end{figure}

Variation of the internuclear distance due to the classical motion of the nuclei are presented in Fig.~\ref{nucdisfor}.
The low intensity ($ 1\times10^{14}$ W cm$^{-2}$) laser field cannot overcome the molecule bonding energy, and thus the internuclear distance does not change considerably and the dissociation process does not take place. 
While, at higher intensities, the field is strong enough to break the bond of the molecule.
It can be seen from Figs.~\ref{H2}-\ref{1zvalue} that the results for the freed and fixed nuclei are very similar.
For the intensity of $5\times10^{14}$, the internuclear distance does not change significantly, up to cycle 6.
During the cycles 6-8, the internuclear distance increases slowly from $\sim3$ to $\sim4.5$ a.u., which leads to a small increase in the populations of H$_{2}^{+}$ and H$_{2}^{+2}$ (thus a decrease in the population of H$_{2}$) for the case with freed nuclei in comparison with the case with fixed nuclei (Fig.~\ref{totalnorm}). 
For the intensity of $ 1\times10^{15}$ W cm$^{-2}$, the internuclear distance does not change significantly up to the end of cycle 4.
During cycles 4-8, the internuclear distance increases from ~2.5 to ~6 a.u. which corresponds to the increased ionization of the H$_{2}$ and H$_{2}^{+}$ (i.e. increased population of H$_{2}^{+}$) when the nuclei are freed (as compared to that of the fixed nuclei).
This effect can be explained by the enhanced ionization (EI) for the internuclear critical distance (R$_{C}$) (\cite{Dehghanian2010,sabzyan2005}). 
For the intensity of $ 5\times10^{15}$ W cm$^{-2}$, H$_{2}$ is ionized mainly up to cycle $\sim3$. 
At this intensity, the predominant population belongs to H$_{2}^{2+}$ which is produced from H$_{2}$ by double ionization and from H$_{2}^{+}$ by single ionization. 
Figure~\ref{nucdisfor} also shows that population of Q-H$_{2}^{2+}$ increases considerably at this intensity.
The internuclear distance does not change significantly during cycles 0-3, but it increases rapidly from $\sim3$ to $\sim8$ a.u. during cycles 3-8, as shown in Fig.~\ref{nucdisfor}. The increase in the internuclear distance results in a small growth of the H$_{2}^{2+}$ population for the case of freed nuclei compared to that of the fixed nuclei, as it is clear in Fig.~\ref{totalnorm}. 

\section*{Total nuclei force on one electron}
 
\begin{figure}
\centerline{\includegraphics[width=15cm]{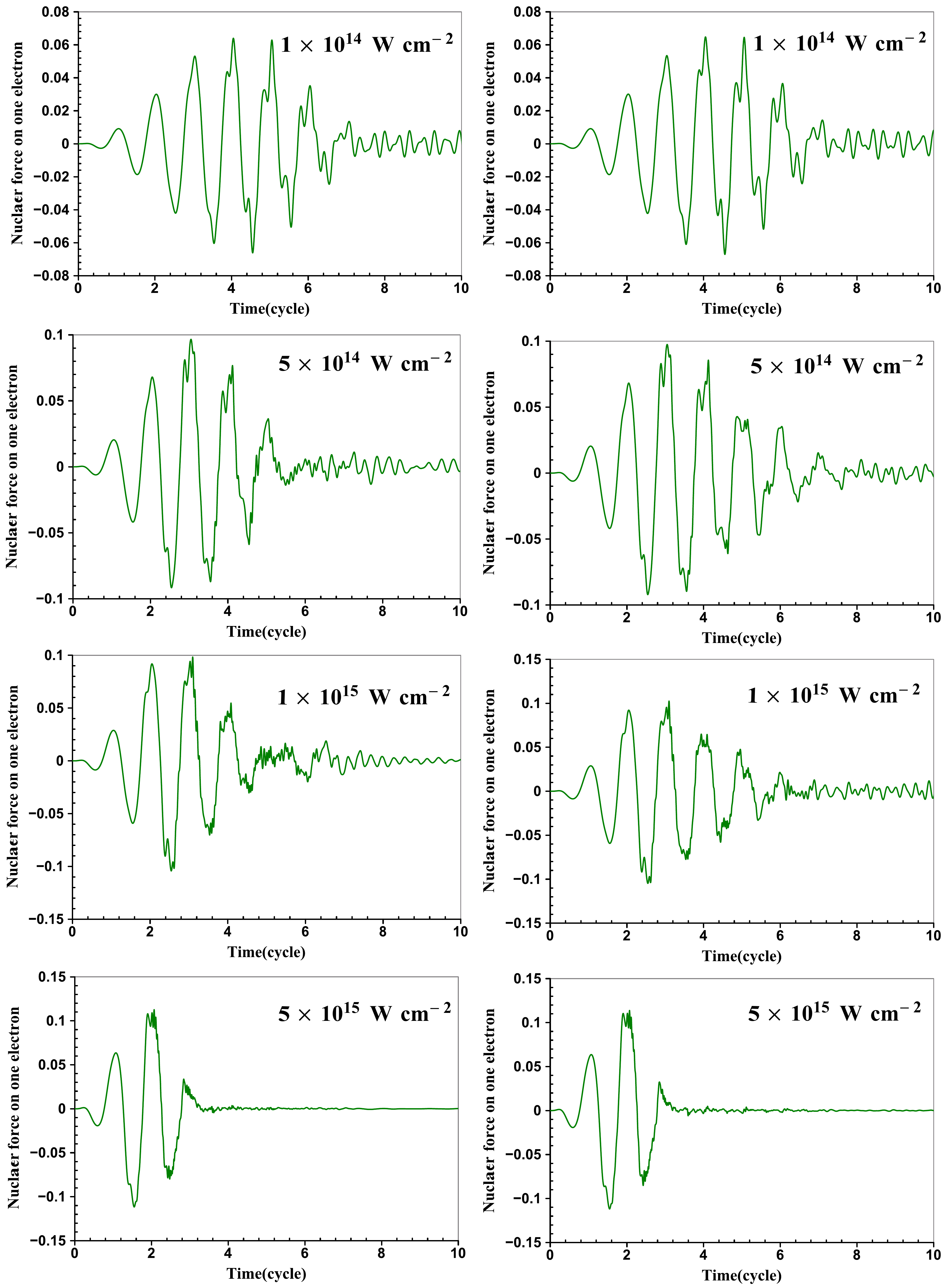}}
\caption{\label{1nucleielectron}
\small (Color online) Variation of the nuclei attractive force on one electron of H$_{2}$ during and after its interaction laser field of different intensities when the nuclei are freed (left) and fixed (right) nuclei.}
\end{figure}

The total force of the nuclei on one of the electrons is extracted using
 \begin{equation}\label{nuclei}
F_{nn-e}=-\Sigma_{\alpha=1}^{2} \int \int \dfrac{[R_{\alpha}(t)-z_{1}] Z_{\alpha} \mid \psi(z_{1},z_{2},t)\mid^{2} }{\lbrace [R_{\alpha}(t)-z_{1}]^{2}+a \rbrace^{3/2}}dz_{1}dz_{2}.
\end{equation}
This force is calculated for different laser intensities and is shown in Fig.~\ref{1nucleielectron}. As it is clear, when the laser field is on, the force is affected by the laser field oscillations so that when the laser pulse is turned off (at the end of cycle 8), the variations of this force becomes negligible. 
At higher intensities, 
the force oscillation  fades out faster even before the end of laser field.
As the laser intensity increases, the oscillation amplitude of the force increases. 
After the termination of the laser field, 
the distribution of electron wave packet on different stationary states causes
a complex oscillation in the expectation value of the nuclei force on one electron, 
similar to what observed for the position expectation value of electrons.

\section*{ Total electron forces on nuclei }

\begin{figure}
\centerline{\includegraphics[width=15cm]{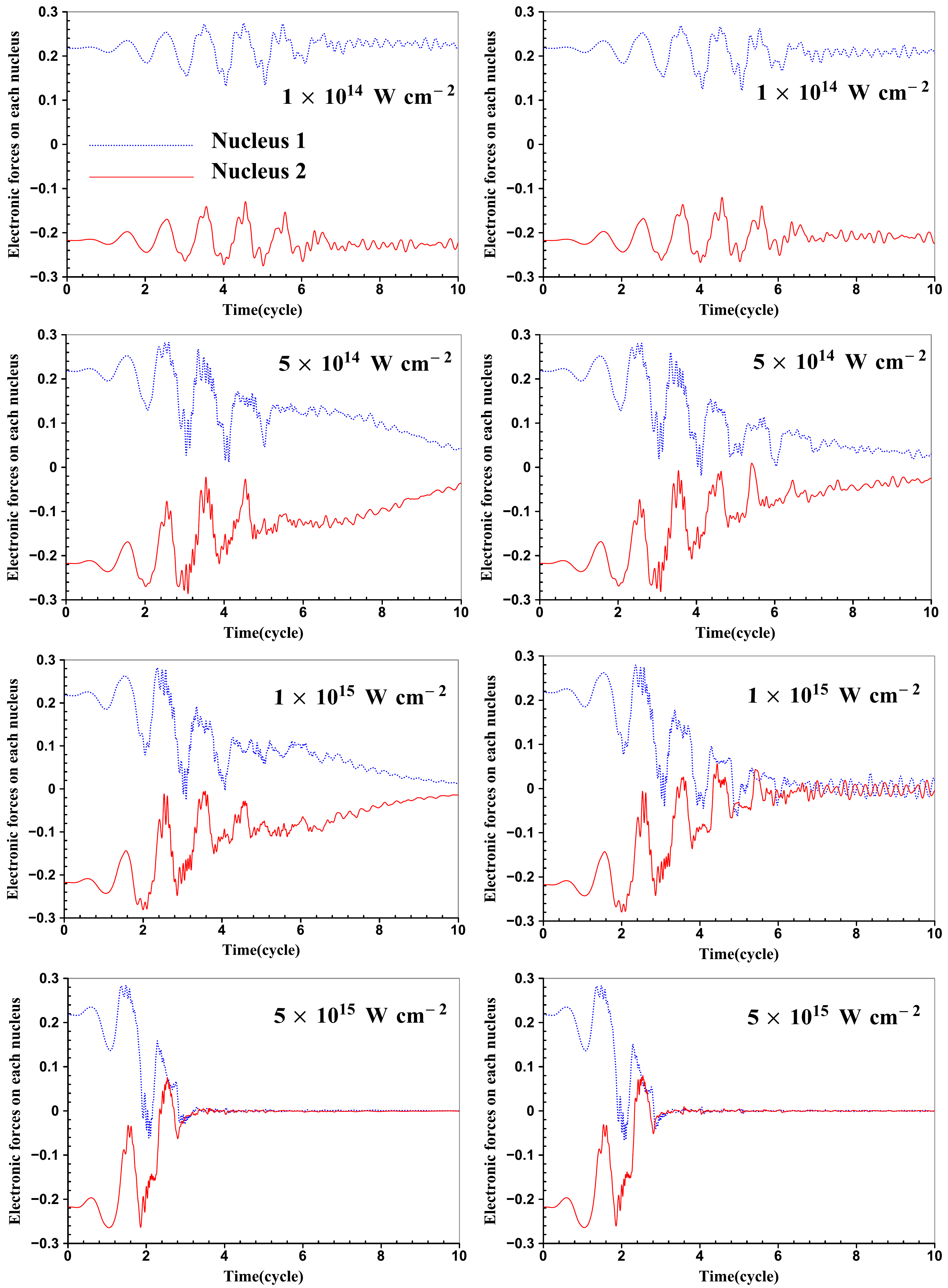}}
\caption{\label{Fee12}
\small (Color online) 
The same as Fig.~\ref{1nucleielectron}, but for the electron forces on the nuclei 1 and 2.
}
\end{figure}

The overall electrons exerted forces applied on the nuclei $\alpha=$ 1 and 2 are obtained using:
\begin{equation}\label{electrons}
F_{ee-n}=-\Sigma_{i=1}^{2} \int \int \dfrac{[R_{\alpha}(t)-z_{i}]\mid \psi(z_{1},z_{2},t)\mid^{2} }{\lbrace [R_{\alpha}(t)-z_{i}]^{2}+a \rbrace^{3/2}}dz_{1}dz_{2}
\end{equation}
The time variation of these forces, F$_{ee-1}$ and F$_{ee-2}$, are shown in Fig.~\ref{Fee12}.
As can be seen in this figure, these forces are affected by the laser field oscillations and approach a constant value after the laser field termination.
At the laser intensity of $ 1\times10^{14}$ W cm$^{-2}$, since no dissociation occurs, these forces are the same for the two cases of moving and fixed nuclei. 
At the intensity of $ 5\times10^{14}$ W cm$^{-2}$, dissociation of the nuclei causes the un-ionized electron wavepackets concentrate around each nuclei and so their effect on the other dissociated nucleus becomes negligible. 
Therefore, when the nuclei are distanced enough or when the electrons are ionized, the force on the nuclei is negligible.
At the intensity of $ 5\times10^{15}$ W cm$^{-2}$, the complete ionization occurs early so that the electron forces become zero at earlier stages of the laser pulse for both cases of the freed and fixed nuclei. 
The differences and similarities of the F$_{ee-1}$ and F$_{ee-2}$, shown in Fig.~\ref{Fee12}, are very interesting. 
In contrast to the electrons, motions of the nuclei are distinguishable. 
The forces on the nuclei are symmetric in the absence and in the early sage of the laser field. 
In the higher $5\times10^{14}$, $1\times10^{15}$, and $5\times10^{15}$ W cm$^{-2}$ intensities, these two forces vary together to follow the variations of the linearly polarized laser pulse.

\section*{Total force on nuclei }

\begin{figure} [tb] 
\centerline{\includegraphics[width=15cm]{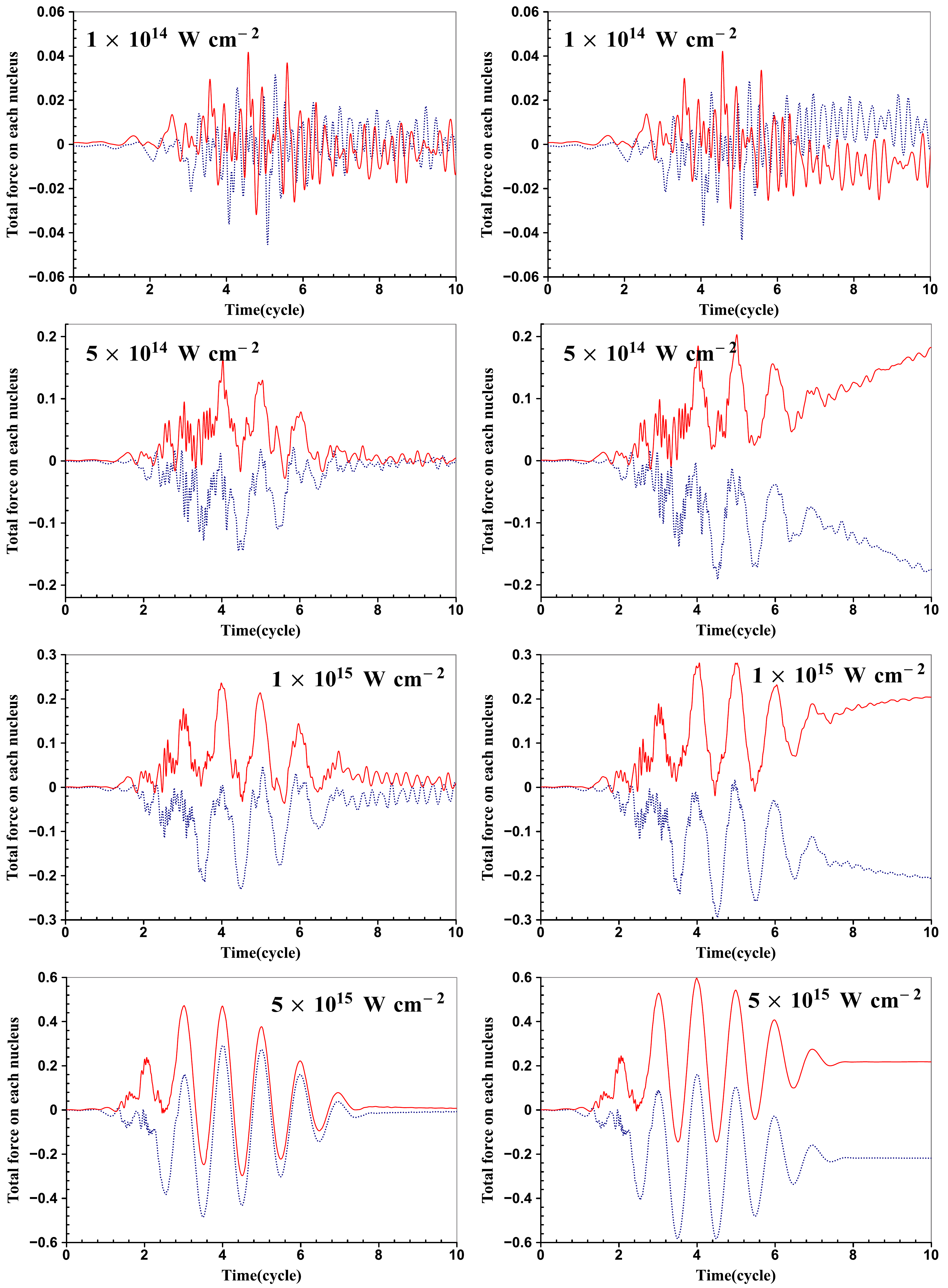}}
\caption{\label{1total}
\small (Color online) 
The same as Fig.~\ref{1nucleielectron}, but for the total force on the nuclei 1 and 2.
}
\end{figure}

The total force which is exerted on the nucleus 1 is shown in Fig.~\ref{1total}. 
In comparison with the dimension of the simulation box, variation of the internuclear distance is very small. 
Therefore, the variations of the forces for both nuclei are very similar and we just report the results obtained for nucleus 1 here. 
In the beginning, the system rests at equilibrium distance. Therefore, the total exerted force on the nucleus 1 is zero. Under irradiation of the laser field and due to the change of the internuclear distance, the net force on nucleus 1 does not vanish.
After turning off the laser pulse, the total force on the nucleus fades out to zero for the freed nuclei. While, in the case of fixed nuclei, the total force never vanishes because of the constant repulsion force of the nuclei.
For the intensity of $ 1\times10^{14}$ W cm$^{-2}$, because of small ionization, the Coulomb attractive and repulsive forces almost cancel each other leading to a negligibly small total force near to zero for both cases of freed and fixed nuclei.

\section*{Repulsive force between the electrons }

\begin{figure} 
\centerline{\includegraphics[width=15cm]{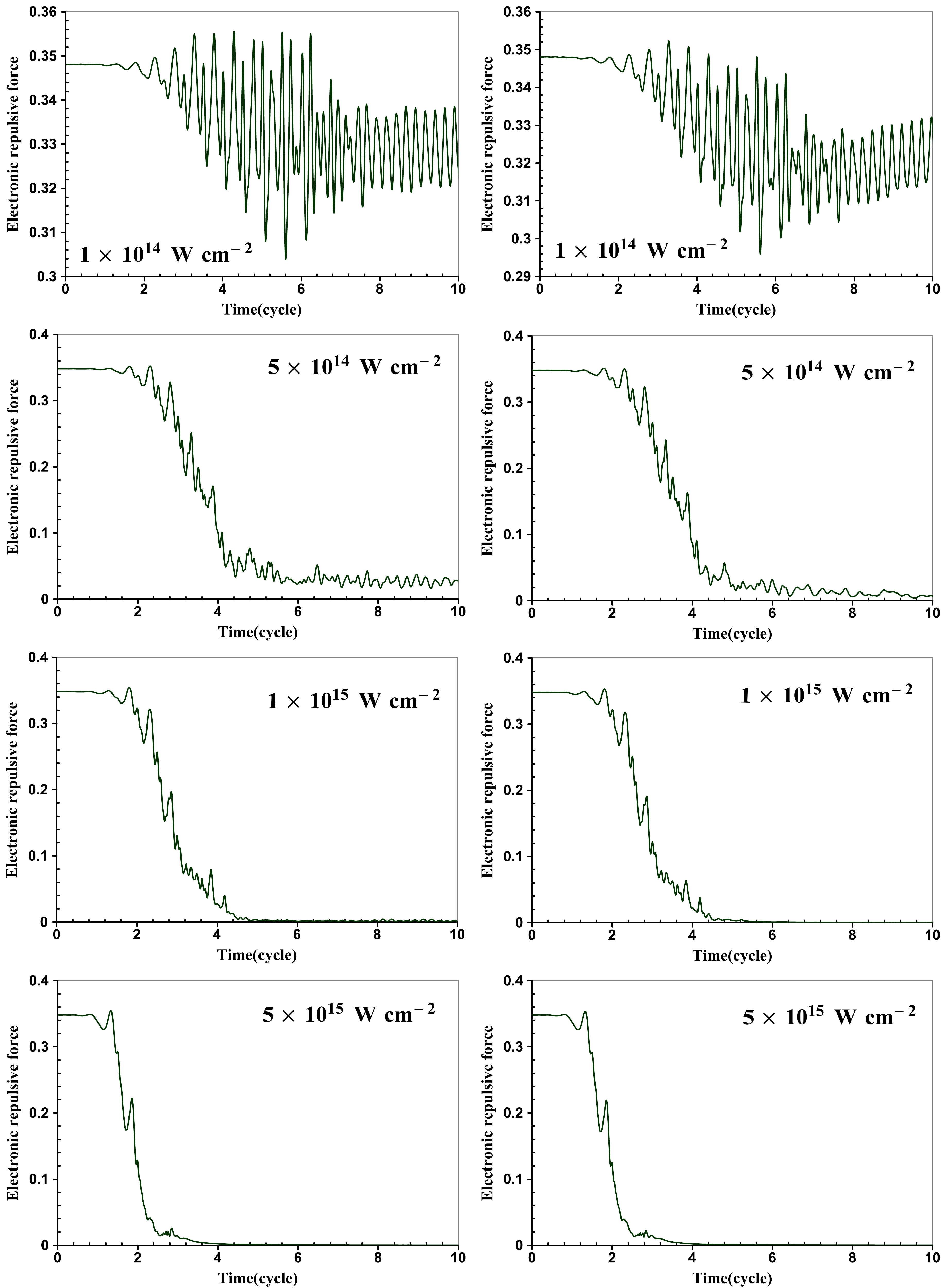}}
\caption{\label{interaction}
\small (Color online) 
The same as Fig.~\ref{1nucleielectron}, but for the repulsive force between the two electrons in the simulation with freed (left) and fixed (right) nuclei.
}
\end{figure}

The expectation value of the repulsive force between the two electrons is calculated by the first order derivative of the repulsive Coulomb potential with respect to the distance between the electrons, i.e. $\langle \dfrac{\partial U_{ee}}{\partial (z_{1}-z_{2})} \rangle$.
Integration is carried out over the entire computational box, and the results are shown in Fig.~\ref{interaction}.
Magnitude of this force reflects, inversely, the distance between the electrons, and therefore, can represent the amount of the correlation between the electrons. 
At the intensity of $ 1\times10^{14}$ W cm$^{-2}$, there is always a noticeable interaction between the electrons
implying that a large part of the correlation between the two electrons is preserved. At higher intensities, repulsive interaction between electrons decreases rapidly, implying that their correlation is decreased effectively 
with increasing intensity of the laser pulse.

\section{Conclusion}
Dynamics of the electrons and the nuclei of hydrogen molecule have been studied, based on the adiabatic approximation, via quantum and classical approaches, respectively by solving time-dependent Schr\"{o}dinger and Newton equations simultaneously.
A one-dimensional model is adopted for both electrons and nuclei coordinates,
and the laser-molecule interaction energy is formulated in the dipole approximation.
Four different intensities have been used; $ 1\times10^{14} $, $ 5\times10^{14} $, $ 1\times10^{15} $ and $ 5\times10^{15} $ W cm$^{-2}$ all with 390 nm wavelength.  
A novel geometry is introduced for the simulation box used in this study. 
In the simulation boxes used in previous works, generally, the dimensions of the $ {H_2}$ and $ {H_2^+}$ ($ d_{H_2}$ and $ d_{H_2^+}$) are assumed to be the same.
This dimension for H$_2$ ($ d_{H_2}$) should be smaller than that for H$_2^+$ ($ d_{H_2^+}$) because the electrons of H$_2$ must escape from a core with one positive charge, while the electron of H$^{+}_{2}$ must escape from a core with two positive charges.
In this work, 
different values are considered for these two dimensions, which allows to reveal more details of the electronic wavepacket evolution and interesting properties of the system.
With introduction of $ d_{H_2^+} > d_{H_2}$, the regions H$_2^+ $(I) and H$_2^+  $(II) overlap in the sections named quasi-H$_2^{+2}$ where both $e_1$ and $e_2$ electrons are ionized.
It is shown that there are three distinct quasi-H$_2^{+2}$ regions.
In addition, the simulation box is designed such that the H$_2^{+}$ components are conserved as long as the intensity of the laser pulse is turned on to assure that the overall second ionization is taken into account properly. 
In this research, evolution of the H$_{2} $ are also detailed by dividing it into four sub-regions related to the ionic H$^+$H$^-$ and covalent (natural) HH states. 
It is shown that when the population of the ionic regions H$^-$H$^+$ (I) and H$^+$H$^-$ (II) becomes higher than that of the covalent regions HH(III), 
ionization causes a considerable reduction in the population of the H$_{2}$ region.
The time-dependent populations of different regions are calculated at different intensities and analyzed comparatively.
It is shown that at the lowest intensity of $ 1\times10^{14}$ W cm$^{-2}$, H$_{2}$ has the dominant population, while at the intensities of $ 5\times10^{14}$ and $ 1\times10^{15}$ W cm$^{-2}$, H$_{2}^{+}$ has the dominant population.
At the highest intensity, $ 5\times10^{15}$ W cm$^{-2}$, the population of the double ionization region, H$_{2}^{2+}$, is dominant.
It is shown that the rapid increase of the double ionization population occurs near the peak of the laser pulse.
The double ionization population of both H$_{2}^{2+}$(I) and H$_{2}^{2+}$(II) regions are the same at the lowest intensity examined here. 
With increasing the laser intensity, the population of the H$_{2}^{2+}$(I) and H$_{2}^{2+}$(II) regions becomes significantly different.
The H$_{2}$ system does not proceed to dissociation at the lowest intensity of $ 1\times10^{14}$ W cm$^{-2}$, but it is dissociated at higher intensities.
Also, the effect of the internuclear distance and motion of nuclei on the enhanced ionization is discussed.
Finally, different time-dependent properties of the system were calculated and analyzed based on the characteristics of the laser pulse and variation of the populations of different regions. 
These properties include 
the internuclear distance, 
the total force of the nuclei on one of the electrons,
the total electron forces on nuclei, 
the total force exerted on the nuclei,
and repulsive force between the electrons.%
\begin{acknowledgments}
%
We thank Dr. A. R. Niknam for providing his computing facilities and Ms. S. Mehrabian and Mr. A. Motevallian for their comments and careful reading of the manuscript. We wish to acknowledge also Shahid Beheshti University for the financial supports, research facilities and High Performance Computing cluster of Laser and Plasma Research Institute(HPCLP).
\end{acknowledgments}
%
\bibliography{p7}

\end{document}